\documentclass[reqno,11pt]{article}  
\newcommand{\version}{Feb.\ 26, 2009}
\usepackage{epsfig,psfrag}
\newlength{\dinwidth}
\newlength{\dinmargin}
\usepackage{amsmath}
\usepackage{amssymb}
\usepackage{latexsym}
\usepackage{cite}
\usepackage{epsfig,psfrag}
\newcommand{\RR}{\mathbb{R}}
\newcommand{\II}{\mathbb{Z}}
\newcommand{\NN}{\mathbb{N}}
\newcommand{\CC}{\mathbb{C}}

\newcommand{\unity}{{\setlength{\unitlength}{1em}
                     \begin{picture}(0.75,1)
                     \put(0,0){$1$}
                     \put(0.34,0){\line(0,1){0.65}}
                     \end{picture}
                   }}
\newcommand{\unit}{{\mbox{\texttt 1}}}

\newcommand{\Ad}{\text{\rm Ad} }
\newenvironment{Proof}%
{\par \medskip \noindent {\em Proof.}}{\hspace*{\fill} $\square$\par%
\medskip\noindent}
\newtheorem{Thm}{Theorem}[section]
\newtheorem{Prop}[Thm]{Proposition} 
\newtheorem{Lem}[Thm]{Lemma} 
\newtheorem{Cor}[Thm]{Corollary} 
\newtheorem{Ass}{Assumption}

\newcommand{\calB}{{\mathcal B}}
\newcommand{\calD}{{\mathcal D}}
\newcommand{\calH}{{\mathcal H}}

\newcommand{\adj}{\dagger}       
\newcommand{\calA}{\mathcal{A}}          
          
\newcommand{\calO}{\mathcal{O}}

\newcommand{\Potild}{\tilde{P}_+^{\uparrow}}
\newcommand{\Potildj}{\tilde{P}_+}
\newcommand{\Po}{P_+^{\uparrow}}
\newcommand{\Lortild}{\tilde L_+^\uparrow}
\newcommand{\Lor}{L_+^{\uparrow}}
\newcommand{\half}{{\frac{1}{2}}} 
\newcommand{\act}{\!\cdot\!}
\newcommand{\clo}{ {\mbox{\bf --}} }
\newcommand{\Wtild}{{\tilde{W}}}

\newcommand{\supp}{{\rm supp}}
\newcommand{\eps}{\varepsilon}
\newcommand{\We}{\Wtild_1}             
\newcommand{\spc}{{C}}             
\newcommand{\Spc}{\mathcal{C}}         
\newcommand{\spcpath}{\tilde{\spc}}    
\newcommand{\Spcpaths}{\tilde{\Spc}} 
\newcommand{\ccc}{{I}}             
\newcommand{\cccpath}{\tilde{\ccc}}    
\newcommand{\Ccc}{\mathcal{K}}         
\newcommand{\Cccpaths}{\tilde{\Ccc}}   
 
\newcommand{\spd}{r}                   
\newcommand{\Spd}{H}              
\renewcommand{\lor}{\lambda}     
\newcommand{\lortild}{\tilde{\lor}} 
\newcommand{\po}{g}              
\newcommand{\potild}{\tilde{\po}}  
\newcommand{\rot}[1]{\tilde{r}(#1)}  
\newcommand{\Rot}[1]{r(#1)}          
\newcommand{\boox}[1]{\tilde{\lambda}_1(#1)}   
\newcommand{\Boox}[1]{\lambda_1(#1)}           
\newcommand{\J}{j}         
\newcommand{\Auni}{\calA} 
\newcommand{\AO}{\calA_0}    
\newcommand{\HO}{\calH_{0}}
\newcommand{\bfp}{{\boldsymbol{p}}}

\newcommand{\Om}{\Omega}
\newcommand{\Out}{{\rm out} }

\newcommand{\lsp}{(\,}
\newcommand{\rsp}{\,)}

\newcommand{\Sec}{\Sigma} 
\newcommand{\SecE}{\Sec^{(1)}} 
\newcommand{\barrho}{{\bar{\rho}}} 
 
\newcommand{\SgeoE}{S_{\rm geo}^{(1)}}  
\newcommand{\SgeoEL}{\hat{S}_{\rm geo}^{(1)}}  
\newcommand{\OmO}{\Omega_0}
\newcommand{\DO}{\Delta_0}  
\newcommand{\JO}{J_0}  
\newcommand{\SO}{S_0}  %
\newcommand{\DF}{\Delta}  
\newcommand{\JF}{J}  
\newcommand{\SF}{S}  %
\newcommand{\SFE}{S^{(1)}}  %
\newcommand{\SFL}{\hat{S}}  
\newcommand{\SFEL}{\hat{S}^{(1)}}  %

\newcommand{\Ue}{U^{(1)}} 
\newcommand{\Uej}{U^{(1)}(j)} 
\newcommand{\UWig}{U^{{\text{\rm w}}}}

\newcommand{\WigRot}{\Omega} 
\newcommand{\UO}{U_0}   
\newcommand{\alphaO}{\alpha}   
\newcommand{\Ee}{E^{(1)}}

\newcommand{\CPTop}{\Theta} 
\newcommand{\F}{{\mathcal F}}  
\newcommand{\FA}{{\mathcal F}_{\rm a}}  
\newcommand{\SA}{S_{\rm a}}  
\newcommand{\JA}{J_{\rm a}}  
\newcommand{\DA}{\Delta_{\rm a}}  
\newcommand{\fa}{{F}_{\rm a}}  

\newcommand{\calS}{{\mathcal S}}
\newcommand{\CE}{\mu}     

\newcommand{\CPT}{CPT }
\newcommand{\Int}{\text{\rm Int}}
\newcommand{\coc}{Z}
\renewcommand{\d}{\text{\rm d}}
\begin{document} 
\title{The  \CPT and Bisognano-Wichmann  Theorems for Anyons and Plektons 
in d=2+1}
\author{Jens Mund\thanks{Supported by FAPEMIG and CNPq.}
\\ 
\scriptsize 
Departamento de F\'{\i}sica, Universidade Federal de Juiz de Fora,
 \\  \scriptsize 
36036-900 Juiz de Fora, MG, Brazil.\\ 
\scriptsize  E-mail: {\tt mund@fisica.ufjf.br}}
\date{ \version\\\vspace{3ex}
{\em 
Dedicated to the memory of Bernd Kuckert.}
}
\maketitle 
\begin{abstract}
We prove the Bisognano-Wichmann and \CPT theorems for massive particles obeying
braid group statistics in three-dimensional Minkowski space. 
We start from first principles of local relativistic quantum theory,
assuming Poincar\'e covariance and asymptotic completeness. 
The particle masses must be isolated points in the mass spectra of the
corresponding charged sectors, and may only be finitely degenerate. 
\end{abstract}
\section*{Introduction} \label{secIntro}
The Bisognano-Wichmann theorem states that a large class
of models in relativistic quantum field theory satisfies 
{modular covariance}, namely: 
The modular unitary group~\cite{BraRob} of the field algebra 
associated to a (Rindler) wedge region coincides with the unitary
group representing the boosts which preserve the wedge.  
Since the boosts associated to all wedge regions generate the
Poincar\'e group, modular covariance implies that the representation
of the Poincar\'e group is  encoded intrinsically in the field algebra. 
This has important consequences, most prominently  
the spin-statistics theorem, the particle/anti-particle
symmetry and the \CPT theorem~\cite{GL95,Kuck}. 
Modular covariance also implies a maximality condition for the field
algebra, namely the duality property~\cite{BiWi}, 
and it implies the Unruh effect~\cite{Unruh}, namely that for a
uniformly accelerated observer the vacuum looks like a heat bath whose
temperature is (acceleration)$/2\pi$. 
The original theorem of Bisognano and Wichmann~\cite{BiWi,BiWi2}
relied on the \CPT theorem~\cite{PauliPCT} and was valid
for finite component Wightman fields. 
However, the physical significance of this latter hypothesis is unclear. 
In the framework of algebraic quantum field theory~\cite{Araki,H96}, 
Guido and Longo have derived modular covariance
in complete generality for conformally covariant theories~\cite{BGL93}. 
In the four-dimensional Poincar\'e covariant case the 
Bisognano-Wichmann theorem has been shown by the author~\cite{M01a} to 
hold under physically transparent conditions, namely for massive theories 
with asymptotic completeness. (Conditions of more technical nature
have been found by several 
authors~\cite{Borchers98,BY00,SHW98,Kuck00,GL00}, see~\cite{BY00} for
a review of these results.)  
In three-dimensional spacetime, however, there may be charged sectors with
braid group statistics~\cite{FM1,F89} containing particles whose 
spin is neither integer nor half-integer, 
which are called Plektons~\cite{FRSII} or, if the statistics is described
by an Abelian representation of the braid group, Anyons~\cite{Wil}.  
In this case, modular covariance and the \CPT theorem are 
also expected~\cite[Assumption~4.1]{FM2} to hold under certain conditions,
but so far have not been proved in a model-independent way. 
The aim of the present article is to prove these theorems from first
principles for massive Poincar\'e covariant Plektons satisfying 
asymptotic completeness.  

Let us comment in more detail on the Bisognano-Wichmann and \CPT 
theorems, and their interrelation, in the familiar case of permutation 
group statistics. Let $W_1$ be the standard wedge 
\begin{equation} \label{eqW1} 
W_1 :=\{\,x\in\RR^3:\,|x^0|<x^1\;\}. 
\end{equation} 
The Tomita operator associated with the field algebra of
$W_1$ and the vacuum is defined as the closed anti-linear 
operator $\SF$ satisfying  
\begin{equation} \label{eqTom}
\SF\,F\Omega :=F^*\Omega, \qquad F\in\F(W_1), 
\end{equation}
where $F^*$ is the operator adjoint and $\F(W_1)$
denotes the algebra of fields localized in $W_1$.  
Denoting its polar decomposition by 
$\SF=J\Delta^{1/2}$, $J$ and $\Delta^{it}$ are called the modular 
conjugation and modular unitary group, respectively, associated with
$\F(W_1)$ and $\Omega$.  
{\em Modular covariance} means that the modular unitary group coincides with 
the unitary
group representing the boosts in $1$-direction (which preserve the
wedge $W_1$), namely:  
\begin{equation} \label{eqModCov'}
  \Delta^{it} = U(\lambda_1(-2\pi t))\,,
\end{equation}
where $\lambda_1(t)$ acts as $\cosh(t)\,\unity+\sinh(t)\,\sigma_1$ on
the coordinates $x^0,x^1$.  
Here, $U$ is the representation  of the universal covering group, 
$\Potild$, of the Poincar\'e group under which the fields are covariant. 
(Note that then, by covariance, the modular groups associated to other
wedges $W=\po W_1$ represent in the 
same way the corresponding boosts $\lambda_W(t)=\po\Boox{t}\po^{-1}$,
and hence the entire representation $U$ is fixed by the modular data.)
The \CPT theorem, on the other hand, asserts the existence of an anti-unitary 
\CPT operator $\Theta$ which re\-presents 
the reflexion\footnote{We consider $\J$ as the $PT$ 
transformation. The total spacetime inversion arises in
four-dimensional spacetime from $\J$ through a $\pi$-rotation about
the $1$-axis, and is thus also a symmetry. In the odd-dimensional case
at hand, $\J$ is the proper candidate for a symmetry (in combination
with charge conjugation), while the total spacetime inversion is
not -- in fact, the latter {\em cannot} be a symmetry in the presence
of braid group statistics~\cite{FM2}\label{ProperCPT}.} 
$\J:=$ diag$(-1,-1,1)$ at the edge of the standard wedge $W_1$ 
in a geometrically correct way:\footnote{In
Eq.~\eqref{eqCPT}, $j\potild j$ 
denotes the unique lift~\cite{Var2} of the adjoint action of $j$ 
from $\Po$ to $\Potild$.}  
\begin{align} \label{eqCPT}
\Theta\,U(\potild)\Theta^{-1} & = U(j\potild j), \quad \potild\in\Potild,\\
\Ad \Theta:\, \F(\spc) & \to \F(j  \spc). \label{eqCPTgeo}
\end{align} 
Here, $\spc$ is a spacetime region within a suitable class. Further, if
a field $F$ carries a certain charge then 
\begin{equation} \label{eqC}
\Theta F \Theta^{-1} \text{ carries the conjugate charge. }
\end{equation}
The \CPT theorem has been used as an input to the proof of modular
covariance by Bisognano and Wichmann~\cite{BiWi,BiWi2}. 
Conversely, the work of Guido and Longo~\cite{GL95}, and
Kuckert~\cite{Kuck}, has shown that modular covariance
implies the \CPT theorem. 
In particular, Guido and Longo have shown~\cite{GL95} that 
modular covariance of the {\em observable} algebra $\calA(W_1)$ 
implies that the
corresponding modular conjugation is a ``PT'' operator on the 
observable level, namely satisfies 
Eq.~\eqref{eqCPT} on the vacuum Hilbert space and Eq.~\eqref{eqCPTgeo}
with $\calA(\spc)$ instead of $\F(\spc)$. 
Further, it intertwines a charged sector with its conjugate sector in
the sense of representations, see Eq.s~\eqref{eqJOImplement} and
\eqref{eqjRhoj} below. 
Therefore, the modular conjugation can be considered a \CPT 
operator. 
These results also hold in theories with braid group statistics. 
In the {\em absence} of braid group statistics, the \CPT theorem can be made
much more explicit~\cite{GL95,Kuck}, namely on the level of the 
field algebra. In fact, in this case the modular conjugation
associated with $\F(W_1)$, multiplied with the 
so-called twist operator, is a \CPT operator in the sense of 
Eq.s~\eqref{eqCPT}, \eqref{eqCPTgeo} and \eqref{eqC}. 

In extending the Bisognano-Wichmann and \CPT theorems to the case of 
braid group statistics, one
encounters several difficulties. Since in this case there are no Wightman
fields\footnote{However there might be, in models, string-localized 
Wightman type fields in the sense of~\cite{St,MSY}.}, the
original proofs of the \CPT and Bisognano-Wichmann theorems do not work.  
Also the proof  of modular covariance in~\cite{M01a} and the
derivation of the explicit \CPT theorem on the 
level of the field algebra from modular covariance~\cite{GL95,Kuck}
do not go through, on two accounts. Firstly, the ``fractional spin''
representations of the universal covering group of the Poincar\'e
group do not share certain analyticity properties of the (half-)
integer spin representations which have been used in~\cite{M01a}. 
This problem has been settled in the article~\cite{M02a}, whose
results have been used to prove the spin-statistics theorem for 
Plektons~\cite{Mu_SpiSta}. 
Secondly, the derivations of modular covariance in~\cite{M01a} and of
the \CPT theorem in~\cite{GL95,Kuck} rely on the existence of 
an algebra $\F$ of charge carrying field operators 
containing the observables $\calA$ as the sub-algebra of invariants 
under a (global) gauge symmetry
and such that the vacuum is cyclic and separating 
for the local field algebras. Such a frame, which we shall call 
the Wick-Wightman-Wigner (WWW) scenario, always exists in the case of
permutation group statistics~\cite{DR90}, but does not exist in the
case of non-Abelian braid group statistics. 
Then it is not even clear what the proper candidate for the 
Tomita operator should be.  We use here a ``pseudo''-Tomita operator 
which has already been proposed by Fredenhagen, Rehren and 
Schroer~\cite{FRSII}. 
A major problem then is that one cannot use the algebraic relations
of the modular objects among themselves and with respect to the field
algebra, and with the representers of the translations. These
relations are asserted by Tomita-Takesaki's and
Borchers'~\cite{Borchers92} theorems, respectively, and enter crucially 
into the derivation of modular covariance and the \CPT
theorem in~\cite{M01a,GL95,Kuck}.  
This problem has been partially settled in~\cite{Mu_BorchersCR}, where 
the algebraic relations of our pseudo-modular objects among themselves 
and with the translations have been analyzed. 

In the present paper, we prove {\em pseudo-modular covariance}, namely
that Eq.~\eqref{eqModCov'} holds with $\Delta^{it}$ standing for the 
pseudo-modular unitary group.  
We also show that the pseudo-modular conjugation 
already is a \CPT operator in the sense of Eq.s~\eqref{eqCPT},
\eqref{eqCPTgeo} and \eqref{eqC}. 
Our line of argument parallels widely that of~\cite{M01a}. 
In the special case of Anyons, there does exist a WWW scenario and we
show modular covariance in the usual sense.  

The article is organized as follows. 
In Section~\ref{secAss} we specify our framework and assumptions in
some detail. 
As our field algebra we shall use the reduced field bundle~\cite{FRSII}. 
This is a $C^*$-algebra $\F$ acting on a Hilbert space which   
contains, apart from the vacuum Hilbert space, subspaces 
corresponding to all charged sectors under consideration. It contains
the observable algebra as the sub-algebra which leaves the vacuum
Hilbert space invariant. However, in contrast to the field algebra in the 
permutation group statistics case it does not fulfil the WWW scenario.  
In particular the vacuum is not separating for the local algebras, and
worse:  {\em Every} field operator $F$ which carries 
non-trivial charge satisfies $F^*\Om=0$. 
Correspondingly, there are no Tomita operators in the literal sense. 
This may be circumvented as proposed in~\cite{FRSII}: There is a 
(non-involutive) pseudo-adjoint $F\mapsto F^\adj$ on $\F$ which 
coincides with the operator adjoint only for observables. 
The point here is that $F\Omega \mapsto F^\adj\Omega$,
$F\in\F(W_1)$, is a well-defined closable anti-linear operator.%
\footnote{To be precise, the ``local'' field  algebras 
depend not only on spacetime regions such as $W_1$, 
but also on certain paths in a sense to be specified in
Section~\ref{secAss}.  In the definition of the pseudo-Tomita 
operator, $W_1$ must therefore be replaced by a path $\We$, 
see Eq.~\eqref{eqSTom}.} 
We define now $S\equiv J\Delta^{1/2}$ as in Eq.~\eqref{eqTom},
with $F^*$ replaced by $F^\adj$, and call $S$, $J$ and $\Delta^{it}$   
the pseudo-Tomita operator, pseudo-modular conjugation and
pseudo-modular unitary group, respectively, associated with 
$\F(W_1)$ --- We add the word
``pseudo'' because $S$ is not a Tomita operator in the strict sense 
(it is not even an involution). 
In Section~\ref{secAlg}, we express the pseudo-Tomita operator $\SF$ in terms
of a family of relative Tomita operators~\cite{Stratila} associated
with the observable algebra $\calA(W_1)$ and certain suitably chosen 
pairs of states, and recall some 
algebraic properties of these objects established in~\cite{Mu_BorchersCR}.  
Using these properties, we show  that the pseudo-modular group 
associated with $\F(W_1)$ leaves this algebra invariant 
(Proposition~\ref{FAInv}), 
just as in the case of a genuine modular group. 
In Section~\ref{secH1} we derive single particle versions of pseudo-modular 
covariance and the \CPT theorem (Corollary~\ref{CorModCov1}) from our 
assumption that the theory be purely massive \eqref{A1}. This 
was already partially implicit in~\cite{Mu_SpiSta}. 
We then prove, in Section~\ref{secHex}, that this property passes over from the
single particle states to scattering states. 
Under the assumption of asymptotic completeness \eqref{A2}, this 
amounts to pseudo-modular covariance of the field algebra. This is our main
result, stated in Theorem~\ref{ModCov}. 
Since the $\adj$-adjoint coincides with the operator adjoint 
on the observables, the restriction of $\SF$ to the vacuum Hilbert
space coincides with the 
(genuine) Tomita operator of the observables.  
We therefore have then modular covariance, in the usual sense, of the
observables. As explained above, this also implies the \CPT theorem on
the level of observables.   
In Section~\ref{secCPT} we make the \CPT theorem explicit and show 
that the pseudo-modular conjugation of the field 
algebra $\F(W_1)$ is a \CPT operator in the sense of 
Eq.s~\eqref{eqCPT}, \eqref{eqCPTgeo}\footnote{In Eq.~\eqref{eqCPTgeo}, 
$\spc$ is now understood to be a path of space-like cones as explained in
Section~\ref{secAss}.} and~\eqref{eqC} (Theorem~\ref{ModCovConj}). 
To this end, we use the mentioned \CPT theorem on the observable 
level~\cite{GL95}, as well as the algebraic properties of the 
``pseudo''-modular objects established in~\cite{Mu_BorchersCR}. 
(The argument used in Section~5 of~\cite{M01a} via scattering theory cannot 
be used since it relies on the fact that the modular conjugation maps the
algebra onto its commutant, which does not hold in the present
case.\footnote{On this occasion, I would like to rectify a minor error in the
argument of~\cite[Section~5]{M01a}. Namely, in Lemma~8 the modular 
conjugation, $J_{W_1}$, must be replaced by $Z^* J_{W_1}$, where $Z$
is the twist operator, and twisted Haag duality for 
wedges~\cite[Eq.~(1.4)]{M01a} must be
assumed. This does not influence the validity of its 
consequences, in particular of the CPT theorem (Proposition 9).})  
In Section~\ref{secAnyons} we finally treat the case of Anyons, where 
there is known to be a 
field algebra $\FA$ in the WWW sense~\cite{MDiss,Re}. In particular, 
the vacuum is cyclic and separating for $\FA(W_1)$, allowing 
for the definition of a (genuine) Tomita operator associated with the wedge. 
Considering the genuine modular objects, we prove modular 
covariance (Theorem~\ref{ModCovAny}).  
We finally show that the modular conjugation, multiplied with an appropriate 
twist operator, is a \CPT operator (Theorem~\ref{CPTAny}). 
This extends the mentioned derivation of the \CPT theorem 
in~\cite{GL95,Kuck} from Bosons and Fermions to Anyons. 
To achieve these results, we exhibit the anyonic field algebra $\FA$ as a
sub-algebra of the reduced field bundle $\F$, and show that the 
corresponding Tomita operator of $\FA(W_1)$ coincides with the 
pseudo-Tomita operator of $\F(W_1)$ (Lemma~\ref{SFFA}). 
\section{General Setting and Assumptions} \label{secAss}
Since we are aiming at model-independent results, we shall use the
general framework of algebraic quantum field theory~\cite{Araki,H96}, where 
only the physical principles of locality, covariance and
stability are required, and formulated in mathematical terms in a quantum
theoretical setting. 
We now specify this setting and make our assumptions precise. 
\paragraph{Observable Algebra.}
The observables measurable in any given bounded spacetime region $\calO$ are
modelled as (the self-adjoint part of) a von Neumann algebra $\AO(\calO)$ of
operators, such that observables localized in causally separated
regions commute.
These operators act in a Hilbert
space $\HO$ which carries a continuous unitary representation 
$\UO$ of the (proper orthochronuous) Poincar\'e group $\Po$ which acts
geometrically correctly: 
\begin{equation} \label{eqCovObs}
\Ad \UO(g):\, \AO(\calO)\to \AO(g\calO),\quad g\in\Po.
\end{equation}
(By $\Ad$ we denote the adjoint action of unitaries.) 
To comply with the principle of stability or positivity of the energy,
the energy-momentum spectrum of $\UO$, namely the joint spectrum of
the generators $P_\mu$ of the spacetime translations,  
is assumed to be contained in the forward light cone. 
The vacuum state corresponds to a unique (up to a factor) 
Poincar\'e invariant vector $\OmO\in\HO$. 
It has the Reeh-Schlieder property, namely it is cyclic
and separating for every $\AO(\calO)$. 
Since the vacuum state should be pure, the net of observables is 
assumed irreducible, $\cap_\calO\AO(\calO)'=\CC\unity$. 
For technical reasons, we also require that the observable algebra
satisfy Haag duality for space-like cones and wedges.\footnote{A 
space-like cone with apex $a$ is a region in Minkowski space of the form 
$\spc=a+\RR^+ \calO$, where $\calO$ is a double cone
whose closure does not contain the origin. A wedge is a region 
which arises by a Poincar\'e transformation from $W_1$, 
see Eq.~\eqref{eqW1}.}   
Namely, denoting by $\Ccc$ the class 
of space-like cones, their causal complements, and wedges, we require  
\begin{equation} \label{eqHD}
\AO(\ccc')= \AO(\ccc)',\quad \ccc \in\Ccc. 
\end{equation}
($\AO(\ccc)$ is defined as the von Neumann algebra generated by all
$\AO(\calO)$, $\calO\subset\ccc$. The prime denotes the causal
complement of a region on the left hand side, and the commutant of an
algebra on the right hand side.) 
For the following discussion of charged sectors, it is convenient to 
enlarge the algebra of observables to 
the so-called universal algebra $\Auni$ generated by isomorphic
images $\Auni(\ccc)$ of the $\AO(\ccc)$, 
$\ccc\in\Ccc$, see~\cite{F90,GL92,FRSII}. 
The family of isomorphisms $\Auni(\ccc)\cong \AO(\ccc)$ extends to a 
representation $\pi_0$ of $\Auni$, the vacuum representation. We then have 
\begin{equation*} 
\AO(\ccc)= \pi_0\Auni(\ccc), 
\end{equation*}
and the vacuum representation is faithful and normal on the local\footnote{We call the algebras
  $\Auni(\ccc)$ ``local'' although the regions $\ccc$ extend to
  infinity in some direction, just in distinction
  from the ``global'' algebra $\Auni$.} algebras
$\Auni(\ccc)$.\footnote{However, $\pi_0$
  is in general not faithful on the global algebra $\Auni$  
due to the  existence of global intertwiners~\cite{FRSII}, see
  Footnote~\ref{GlobIn}.} 
The adjoint action~\eqref{eqCovObs} of the Poincar\'e group on the
  local algebras lifts to a representation by automorphisms
  $\alphaO_\po$ of $\Auni$, $g\in\Po$, which acts geometrically correctly: 
\begin{align*} 
\Ad \UO(\po) \circ \pi_0 &= \pi_0\circ \alphaO_\po,\\
\alphaO_\po:\;  \Auni(\ccc)&\to   \Auni(\po \ccc). 
\end{align*}
\paragraph{Charged Sectors.}
A superselection sector is an equivalence class of irreducible 
representations $\pi$ of the algebra $\AO$ of quasi-local observables,
namely the $C^*$-algebra generated by all local 
observable algebras $\AO(\calO)$. 
As a consequence of our  Assumption~\ref{A1}, we shall deal only with 
representations which are localizable in space-like cones~\cite{BuF},
i.e., equivalent to the vacuum
representation when restricted to the causal complement of any space-like
cone. 
Such representation uniquely lifts to a representation of the universal algebra
$\Auni$. If Haag duality~\eqref{eqHD} holds, it is
equivalent~\cite{DHRIII,F90} to a
representation of the form $\pi_0\circ \rho$ acting in $\HO$, 
where $\rho$ is an endomorphism of $\Auni$ {localized} in some 
specific region $\spc_0\in\Ccc$ in the sense that 
\begin{equation} \label{eqRhoInC0}
\rho(A)=A\qquad \text{ if } \; A \in\Auni(C_0'). 
\end{equation} 
The endomorphism $\rho$ is further {transportable} to other 
space-like cones, which means that 
for every space-like cone $\spc_1$ and $\ccc\in\Ccc$ containing
both $\spc_0$ and $\spc_1$, there is a unitary $U\in\Auni(I)$ such that 
$\Ad U\circ \rho$ is localized in
$\spc_1$.\footnote{In $2+1$ dimensions, for every pair of causally 
separated space-like cones
  $\spc_0,\spc_1$ there are {two} topologically distinct ways to
  choose $\ccc\supset\spc_0\cup\spc_1$: Either one has to go clockwise
from $\spc_0$ to $\spc_1$ within $\ccc$, or anti-clockwise. This is
the reason for the existence of global self-intertwiners in $\Auni$
which are in the kernel of the vacuum representation, and makes the
enlargement from $\AO$ to $\Auni$ necessary. It is also the
reason for the occurrence of braid group statistics in $2+1$
dimensions. \label{GlobIn}}
We shall call localized and transportable endomorphisms simply
{\em localized morphisms}. 
We further assume the representation $\pi\cong \pi_0\rho$ to be
covariant with positive energy. That means that there is a 
unitary representation 
$U_\rho$ of the universal covering group $\Potild$ of the Poincar\'e
group with spectrum contained in the forward light cone 
such that 
\begin{equation} \label{eqCovRho} 
\Ad U_\rho(\potild) \circ \pi_0\rho = 
\pi_0\rho \circ \alphaO_\po, \quad \po\in\Po,
\end{equation}
where $\potild$ is any element of $\Potild$ mapped onto $\po$ 
by the covering projection. 
Superselection sectors, namely equivalence classes of localizable  
representations of $\AO$, 
are in one-to-one correspondence with inner equivalence
classes of localized morphisms of $\Auni$. 
They are the objects of a category whose three crucial
structural elements are products, conjugation and sub-representations.  
More specifically, {\em products} 
$\rho_1\rho_2:=\rho_1\circ\rho_2$ 
of localized morphisms are again localized morphisms, leading to 
a composition of the corresponding sectors. 
A morphisms $\rho$ localized in $\spc$ is said to contain another 
such morphism $\tau$ as a {\em sub-representation} if there is a
non-zero observable $T\in\Auni$, such that 
\begin{equation*} 
\rho(A)\,T  = T \, \tau(A) \quad \text{ for all } A\in\Auni.
\end{equation*}
(If both $\rho$ and $\tau$ are localized in a space-like cone $\spc$,
then $T\in\Auni(\spc)$ by Haag duality.) 
An observable $T$ satisfying this relation is called an intertwiner
from $\tau$ to $\rho$. The set of all such intertwiners 
is denoted as $\Int(\rho|\tau)$. They are the arrows between the
objects $\rho$ and $\tau$. Arrows can be composed if
they fit together and have adjoints.  Namely:   
If $T\in\Int(\rho|\tau)$ and $S\in\Int(\tau|\sigma)$ then $T\circ
S\in\Int(\rho|\sigma)$; if $T\in\Int(\rho|\tau)$ then 
$T^*\in\Int(\tau|\rho)$. It follows that if $\tau$ is irreducible 
(i.e., the representation $\pi_0\circ\tau$ of $\Auni$ is irreducible), then 
$\Int(\rho|\tau)$ is a Hilbert space with scalar product $\langle T,S
\rangle$ being fixed by 
\begin{equation*} 
\langle T,S \rangle\,\unity := \pi_0(T^*S),\quad T,S\in\Int(\rho|\tau).
\end{equation*} 
There is also a product on the arrows, namely: 
if $T\in\Int(\rho|\hat\rho)$ and $S\in\;\Int(\sigma|\hat\sigma)$ then 
\begin{equation} \label{eqTxS}
T\times S \, := T\,\hat\rho(S)\equiv \rho(S)\,T \quad \in
\Int(\rho\sigma|\hat\rho\hat\sigma). 
\end{equation} 

As a consequence of our  Assumption~\ref{A1}, all morphisms
considered here have finite statistics~\cite{F81},  i.e.\ 
the so-called  
statistics parameter $\lambda_\rho$~\cite{DHRIII} is non-zero. 
This implies~\cite{DHRIV} the existence of a 
{\em conjugate} morphism $\bar\rho$ 
characterized, up to equivalence, by the fact that the composite 
sector $\pi_0\bar\rho\rho$ 
contains the vacuum representation $\pi_0$ precisely once.
Thus there is a unique, up to a factor, intertwiner $R_\rho\in\Auni(\spc_0)$
satisfying $\barrho\rho(A) R_\rho = R_\rho A$ for all $A\in\Auni$.  
The conjugate $\barrho$ shares with $\rho$ the properties of 
covariance~\eqref{eqCovRho}, finite statistics, and
localization~\eqref{eqRhoInC0} in
some space-like cone which we choose to be $\spc_0$. 
Using the normalization convention of \cite[Eq.~(3.14)]{DHRIV}, 
namely $R_\rho^*R_\rho=|\lambda_\rho|^{-1}\unity$, 
the positive linear endomorphism $\phi_\rho$ of $\Auni$  defined as
\begin{equation} \label{eqLeftInvR}
\phi_\rho(A)= |\lambda_\rho|\; 
R_\rho^*\bar\rho(A) R_\rho 
\end{equation}
is the unique left inverse~\cite{DHRIV,BuF} of $\rho$. 
In the present situation of three-dimensional space-time, 
the statistics parameter 
$\lambda_\rho$ may be a complex non-real number, 
corresponding to braid group statistics. We admit the case when its 
modulus is different from one (namely when $\rho$ is not surjective), 
corresponding to non-Abelian braid group statistics. 
\paragraph{Field Algebra.} 
From the observable algebra and a set of relevant sectors a field 
algebra can be constructed  in various ways, see for example 
\cite{FM2,FRSII,MackSchom,FroKer,Fuchs95,Rehren96}. Some of these
constructions have a quantum group or a more general structure 
playing the role of a global gauge group, however none of them fulfils
the WWW scenario. Unfortunately, most constructions work only for 
models with a finite set of charges, whereas we wish to
consider here an arbitrary (though countable) number of charges. 
We choose as field algebra the reduced field bundle proposed 
in~\cite{FRSII}, which in turn has been based
on the field bundle of~\cite{DHRIV}. 
We start with a countable collection $\Sec$   
of pairwise inequivalent localized, covariant, irreducible morphisms with
finite statistics, one from each relevant sector, which is stable under 
conjugations and composition with subsequent reduction, and contains
the identity morphism $\iota$. We take all morphisms to be 
localized in the same space-like cone $\spc_0$, which we choose to
be contained in $W_1$. 
The space of all relevant {states} is described by the 
Hilbert space 
\begin{equation} \label{eqHS} 
\calH:= \bigoplus_{\rho\in\Sec} \calH_\rho, \quad
\calH_\rho =  \HO. 
\end{equation}
We shall denote elements of $\calH_\rho$ as $(\rho,\psi)$. 
For each $\rho\in\Sec$ there is a unitary representation $U_\rho$ 
of $\Potild$ acting in $\HO$. 
(For $\rho=\iota$, we take $U_\iota\equiv U_0$.) 
This gives rise to the direct sum representation $U$ on $\calH$ 
\begin{equation} \label{eqU} 
U(\potild) \,(\rho,\psi) :=(\rho,U_\rho(\potild)\psi).
\end{equation}
The vacuum vector 
\begin{equation*} 
\Omega:= (\iota,\OmO) \;\in\calH_\iota  
\end{equation*}
is invariant under this representation. 
The observables act in $\calH$ via the direct sum of all relevant
representations $\pi_0\circ\rho=:\pi_0\rho$, 
\begin{equation*} 
\pi(A) \,(\rho,\psi):= (\rho,\pi_0\rho(A)\psi). 
\end{equation*}
The idea of a charge carrying field is that it should add a certain charge 
$\rho_c$ to a given state $\psi\in\calH_{\rho_s}$. But since
the product morphism $\rho_s\rho_c$ is, in general, not contained in the chosen
set of irreducible morphisms, the new state must be projected onto an
irreducible sub-representation $\rho_r\in\Sec$ of
$\rho_s\rho_c$. (The subscripts $s,c,r$ stand for ``source'', ``charge''
and ``range'', respectively.) 
This idea is realized as follows~\cite{FRSI,FRSII}. 
Given any three $\rho_s,\rho_c,\rho_r\in\Sec$ such that $\rho_s\rho_c$ contains
$\rho_r$ as a sub-representation, the corresponding intertwiner space 
$\Int(\rho_s\rho_c|\rho_r)$ has a certain finite~\cite{FRSII} dimension
$N$. We choose an orthonormal basis, i.e., a collection 
$T_i\in\Auni(\spc_0)$, $i=1,\ldots,N$, satisfying 
\begin{equation*} 
T_i^*\,T_j = \delta_{ij}\unity,
\quad \sum_{i=1}^N T_i\,T_i^* =\unity_{\rho_s\rho_c},  
\end{equation*}
where $\unity_{\rho_s\rho_c}$ is the unit in the algebra 
$\Int(\rho_s\rho_c|\rho_s\rho_c)$. 
Following~\cite{Re90b}, we shall call the multi-index 
\begin{equation*} 
e:= (\rho_s,\rho_c,\rho_r,i)
\end{equation*}
a ``superselection channel'' of type $(\rho_s,\rho_c,\rho_r)$, and
denote any one of the $T_i$ from above generically as 
$$
T_e\;\in\Int(\rho_s\rho_c|\rho_r).
$$
We shall also call $s(e):=\rho_s$, $c(e):=\rho_c$ and $r(e):=\rho_r$ the
source, charge and range of $e$, respectively. 
If $s(e)$ or $c(e)=\iota$, we choose $T_e=\unity$. 
The charge carrying fields are now defined as follows. Given $e$ of
type $(\rho_s,\rho_c,\rho_r)$ and $A\in\Auni$, $F(e,A)$ is 
the operator in $\calH$ defined by 
\begin{equation*} 
F(e,A)\,(\rho,\psi):= \delta_{\rho_s,\rho}\; 
\big(\rho_r,\pi_0\big(T_e^*\rho(A)\big)\psi\big). 
\end{equation*}
Heuristically, this describes the action of $A$ in the background charge
$\rho$, addition of the charge $\rho_c$ and subsequent projection onto
$\calH_{\rho_r}$ via the intertwiner $T_e^*$. 
The norm-closed linear span of all these operators,  
\begin{equation*} 
\F:= \big(\bigoplus_{e} \{ F(e,A),\; A\in\Auni\,\}\big)^\clo,
\end{equation*}
where the sum goes over all superselection channels $e$, is closed 
under multiplication and will be called the field algebra. It
is in fact a $C^*$ sub-algebra of $\calB(\calH)$~\cite{FRSI}. 
It contains the (representation $\pi$ of the)
observable algebra, namely 
$$
\pi(A) =  \sum_{e:\,c(e)=\iota} F(e,A).  
$$
\paragraph{{\it Localization.}} 
Fields are {localizable} to the same extent to which the charges are
localizable which they carry, namely in unbounded regions in the class $\Ccc$. 
In order to have any definite space-like commutation relations, the
{fields} need to carry some supplementary information in addition to
the localization region, due to the existence of global intertwiners (see
Footnote~\ref{GlobIn}).  The possibility we choose is to consider
paths in $\Ccc$ starting from our fixed reference 
cone $\spc_0$.\footnote{Two other possibilities are: 
To introduce a reference space-like cone from
which all allowed localization cones have to keep space-like separated 
(this cone playing the role of a ``cut'' in the
context of multi-valued functions)~\cite{BuF}; or a cohomology theory of 
nets of operator algebras as introduced by 
Roberts~\cite{Roberts,Roberts76,Roberts80}.} 
 By a {\em path} in $\Ccc$ from $\spc_0$ to a region $\ccc\in\Ccc$ (or
 ``ending at'' $\ccc$) we mean a finite
sequence $(I_0,\ldots,I_n)$ of regions in $\Ccc$ with $I_0=\spc_0$,
$I_n=\ccc$, such that either $I_{k-1}\subset I_k$ or
$I_{k-1}\supset I_k$ for $k=1,\ldots,n$. 
Given a path $(\spc_0=I_0,I_1,\ldots,I_n=\ccc)$ and a morphism 
$\rho\in\Sec$ there are unitaries $U_k\in\Auni(I_{k-1}\cup I_k)$ such that
$\rho_k:=\Ad(U_k\cdots U_1)\circ \rho$ is localized in $I_k$. We shall
call $U:=U_n\cdots U_1$ a {\em charge transporter} for $\rho$ along the path
$(I_0,\ldots,I_n)$.
Now a field operator $F(e,A)$ with $c(e)=\rho$, is said to be
{\em localized} along a path in $\Ccc$ ending at $\ccc$ 
if there is a
charge transporter $U$ for $\rho$ along the path such that 
$$
UA\in\Auni(\ccc). 
$$
This localization concept clearly depends only on the homotopy classes
(in an obvious sense~\cite{FRSII}) of paths. 
We shall denote the homotopy class of a path
ending at $\ccc$ by $\cccpath$, and the set of all such (classes of) 
paths by $\Cccpaths$.  
The field operators localized in a given path $\cccpath$ generate a
sub-algebra of $\F$ which we denote by $\F(\cccpath)$. 
The vacuum $\Omega$ is cyclic for the local fields,
i.e.\ for any path $\cccpath$ there holds 
\begin{equation} \label{eqVacCyc}
\big(\F(\cccpath)\,\Omega\big)^\clo = \calH. 
\end{equation}
Note, however, that $\Omega$ is not separating for the 
local\footnote{Again, we call the algebras 
$\F(\cccpath)$ ``local''  just in distinction to the ``global''
algebra $\F$.} field 
algebras $\F(\cccpath)$, 
since every field with non-trivial source 
annihilates the vacuum.  

We now give an alternative description of $\Cccpaths$ in the spirit of the
``string-localized'' quantum fields proposed in~\cite{MSY}, which  will be 
useful in the sequel. 
It is based on the observation that  
a space-like cone $\spc$ is characterized by its apex
$a\in\RR^3$ and the space-like directions contained in $\spc$.  
Namely, let $\Spd$ be the manifold of space-like directions, 
\begin{equation*} 
\Spd :=\{ \spd\in\RR^3,\; \spd\cdot \spd=-1\}. 
\end{equation*} 
The set of space-like directions contained in $\spc$ is 
$\spc^\Spd:=(\spc-a)\cap \Spd$, and there holds $\spc=a+ \RR^+ \spc^\Spd$.  
$\spc^\Spd$ is in fact a double cone in $\Spd$.\footnote{This is so
  because the boundary of $\spc-a$ consists of 4 (pieces of) light-like
  planes through the origin. The intersection of such a
  plane with $\Spd$ is a light-like geodesic in
  $\Spd$~\cite[proof of Prop.~28]{ONeill}. Thus, $\spc^\Spd$ is bounded
by 4 light-like geodesics emanating from two time-like separated
points, and therefore is a double cone in the
two-dimensional spacetime $\Spd$.}
A similar consideration holds for causal complements of space-like
cones and wedge regions (except that the apex of a wedge is fixed only
modulo translations along its edge). 
We can therefore identify regions in $\Ccc$ with regions of the form 
\begin{equation} \label{eqRH}
\{a\}\times \ccc^\Spd\quad\subset\quad \RR^3\times \Spd,
\end{equation}
where $\ccc^\Spd$ is a double cone, a causal complement thereof, or a
wedge,  in $\Spd$. 
Let us denote the class of such regions by $\Ccc^\Spd$. 
Regions in $\Ccc^\Spd$ are simply connected, whereas $\Spd$ itself has 
fundamental group $\II$. 
Thus the portion of the universal covering space of $\Spd$ over a 
region $\ccc^\Spd$ in $\Ccc^\Spd$ consists of a 
countable infinity of copies (``sheets'') of $\ccc^\Spd$. 
We shall generically denote such a sheet over $\ccc^\Spd$ by
$\cccpath^\Spd$, and we denote by $\Cccpaths^\Spd$ the class of such
sheets. 
We identify the universal covering space $\tilde\Spd$ of 
$\Spd$ with homotopy classes of
paths in $\Spd$ starting at some fixed reference direction $\spd_0$, which
we assume to be contained in the reference cone $\spc_0$. (A sheet   
$\cccpath^\Spd$ is canonically homeomorphic to $\ccc^\Spd$, but contains in
addition the information of a winding number distinguishing it
from the other sheets over $\ccc^\Spd$, see Figure~1.) 
We now identify paths $\cccpath\in\Cccpaths$ with regions of the form 
\begin{equation} \label{eqRHTilde}
\{a\}\times \cccpath^\Spd\quad\subset\quad \RR^3\times \tilde\Spd, 
\end{equation}
as follows. Given
a path $(\ccc_0=\spc_0,\ccc_1,\ldots,\ccc_n=\ccc)$ in $\Ccc$, pick a path
$\gamma=\gamma_n\ast\cdots \ast\gamma_0$ in $\Spd$ from 
$\spd_0$ to some $\spd$ contained in $\ccc$, and points 
$a_0,\ldots,a_n$ in $\RR^3$ with $a_0=0$ and $a_n=a=$ apex of $\ccc$, such that
$\gamma_k(t)\in\ccc_k-a_k$ for $t\in[0,1]$, $k=0,\ldots,n$. Then we
associate with $(\ccc_0,\ldots,\ccc_n)$ the region~\eqref{eqRHTilde}, 
where $\cccpath^\Spd$ is the unique sheet over
$\ccc^\Spd=(\ccc-a)\cap \Spd$ which contains the homotopy class of $\gamma$.  
Different paths $\gamma$ lead to the same sheet, and the sheet depends
only on the ``homotopy class'' (in the sense of~\cite{FRSII}) of
$(\ccc_0,\ldots,\ccc_n)$.  
Therefore the above prescription defines a one-to-one correspondence
between $\Cccpaths$ and $\RR^3\times\Cccpaths^\Spd$, which shall be used to
identify them. 
In this identification, the covering space aspect of $\Cccpaths$ shows
up in ``accumulated angles'', endowing $\Cccpaths$ with a partial order
relation.  
Namely, given $\cccpath_i=\{a_i\}\times \cccpath_i^\Spd$ with $\ccc_1$ and 
$\ccc_2$ space-like separated, we shall write 
\begin{equation*} 
\cccpath_1<\cccpath_2
\end{equation*}
if for any $[\gamma_i]\in \cccpath_i^\Spd$ there holds 
$\int_{\gamma_1}d\theta<\int_{\gamma_2} d\theta$, where $d\theta$
denotes the angle one-form in a fixed Lorentz frame. (This is
well-defined since the last relation is independent of the
representants of $[\gamma_i]$ and of the Lorentz frame.)   
\begin{figure}[ht] 
 \label{Fig1}
\psfrag{e0}{$\spd_0$}
\psfrag{e1}{$\gamma_1$}
\psfrag{e2}{$\gamma_2$}
\psfrag{e3}{$\gamma_3$}
\psfrag{C}{$C^H$}
\psfrag{H}{$H$}
\begin{center}
\epsfxsize32ex 
\epsfbox{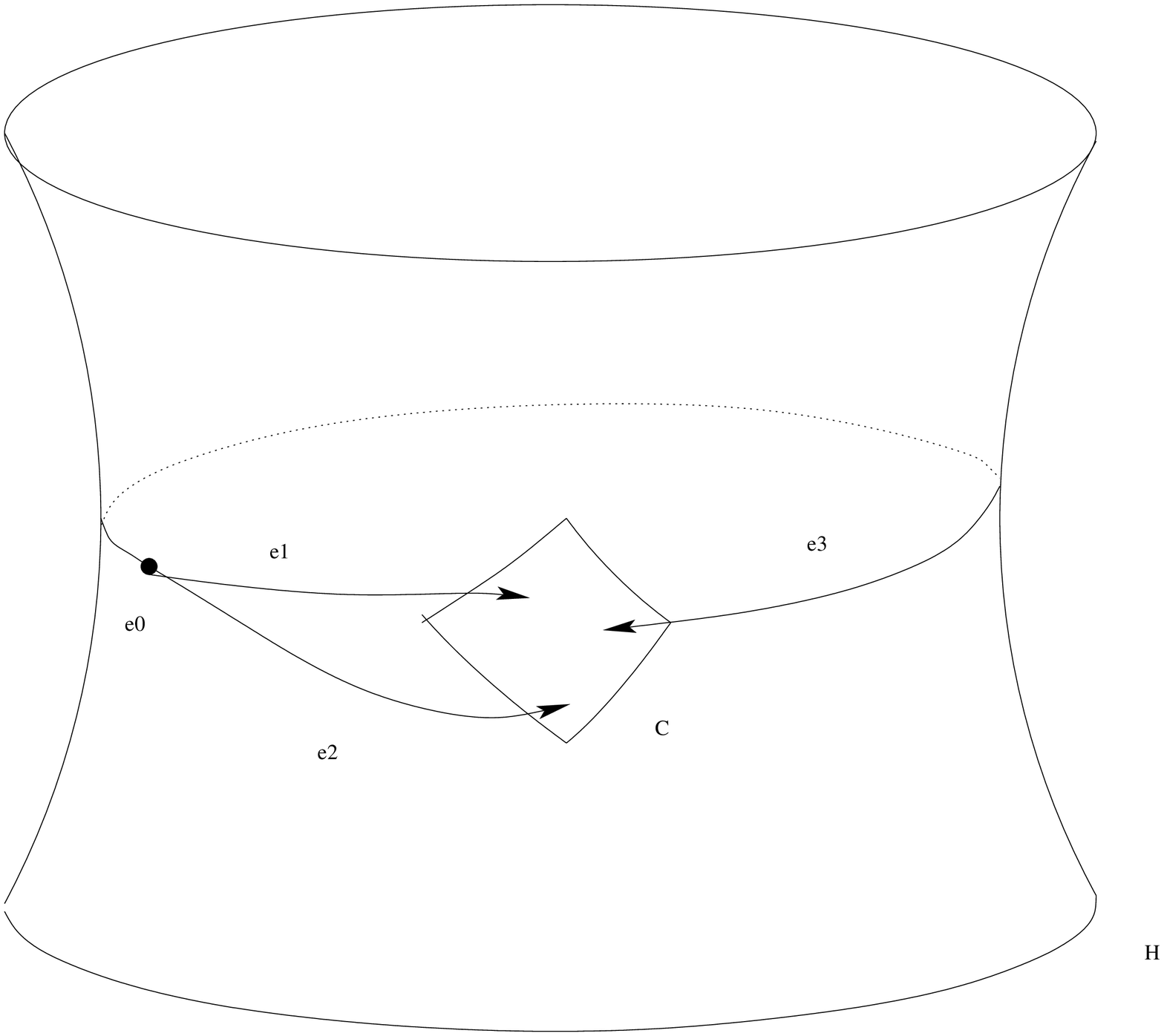}
\caption{The (classes of the) paths $\gamma_1$ and $\gamma_2$ lye in the same
sheet, say $\spcpath^\Spd$, over $\spc^H$. The path $\gamma_3$ lies in
a different
sheet, which is mapped by a $2\pi$ rotation onto $\spcpath^\Spd$. }
\end{center}
\end{figure}
\paragraph{{\it Covariance.}} 
The adjoint action of the representation $U$ of $\Potild$ leaves the
field algebra invariant, more specifically~\cite{DHRIV}: 
\begin{equation} \label{eqUImplement} 
U(\potild)\,F(e,A) U(\potild)^{*} 
= (e, Y_\rho(\potild)\,\alpha_\po(A)),
\end{equation}
where $\rho=s(e)$ is the charge of $e$, and the so-called cocycle 
$Y_\rho(\potild)\in\Auni$ is characterized by 
\begin{equation} \label{eqYrho} 
\pi_0\big(Y_\rho(\potild)\big) = U_\rho(\potild)U_0(\po)^*.
\end{equation}
The adjoint action on the fields is geometrically correct, i.e., 
\begin{equation*} 
\Ad U(\potild):\F(\cccpath)\to\F(\potild\act\cccpath). 
\end{equation*}
Here, $\potild\cdot\cccpath$ denotes the natural action of the 
universal covering of the Poincar\'e group on $\Cccpaths$, defined as follows. 
Let $\potild=(x,\lortild)$, where 
$x$ is a spacetime translation and $\lortild$ is an element of the
universal covering group $\Lortild$ of the Lorentz group, projecting onto
$\lor\in\Lor$. Then 
\begin{equation} \label{eqPoincSpc}
(x,\lortild)\act \big(\{a\}\times \cccpath^\Spd\big):= 
\{x+\lor a\}\times \lortild \act\cccpath^\Spd, 
\end{equation}
where $\lortild\act\cccpath^\Spd$ denotes the lift of the action of 
the Lorentz group on $\Spd$ to the respective universal covering spaces. 
The rotations about integer multiples of $2\pi$ do not act
trivially, but rather coincide with the action of the fundamental group,
$\II$, on the universal covering space of $\Spd$. Namely, they change
winding numbers, see Figure~1. Related to this, we  
define the {\em relative winding number} $N(\cccpath_2,\cccpath_1)$ of
$\cccpath_2$ w.r.t.\ $\cccpath_1$ to be the unique integer $n$ such that 
\begin{equation*} 
\rot{2\pi n}\act \cccpath_1 < \cccpath_2 < 
\rot{2\pi (n+1)} \act \cccpath_1, 
\end{equation*}
where $\rot{\cdot}$ denotes the rotation subgroup in 
$\Lortild$. See Fig.~2 for an example. 
(Note that this number is independent of the choice of reference 
direction $\spd_0$.)  
\begin{figure}[ht] 
\psfrag{e0}{$\spd_0$}
\psfrag{C1}{$\spcpath_1^\Spd$}
\psfrag{C2}{$\spcpath_2^\Spd$}
\psfrag{H}{$H$}
\begin{center}
\epsfxsize30ex 
\epsfbox{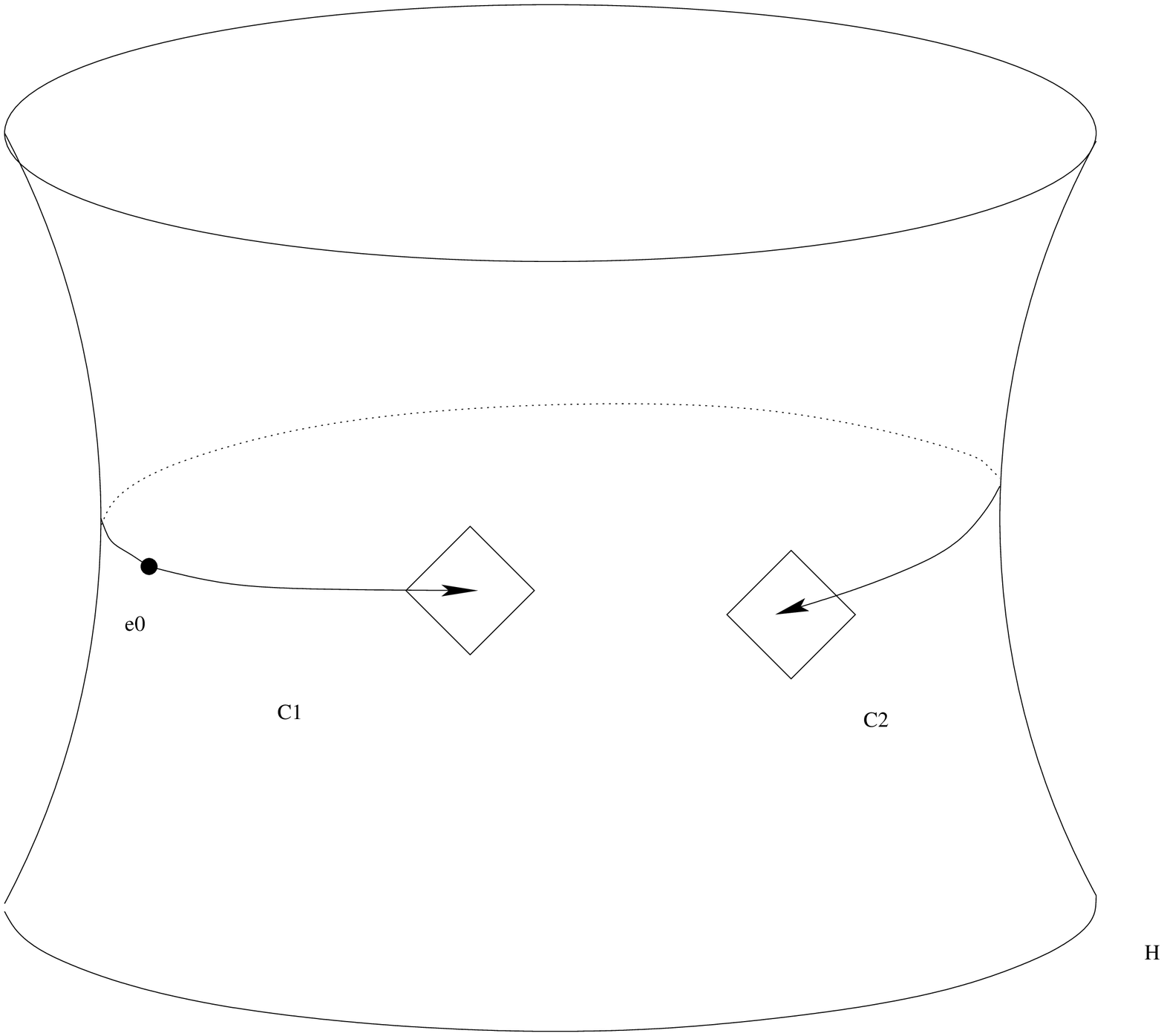}
\caption{$\spcpath_1$ and $\spcpath_2$ have relative winding number 
$N(\spcpath_2,\spcpath_1)=-1$.} 
\end{center}
\end{figure}
\paragraph{{\it Pseudo-Adjoint.}} 
Let $\F_\iota$ be the Banach space generated by field operators
$F(e,A)\in\F$ which have trivial source, $s(e)=\iota$, i.e., 
which have $e$ of the form $(\iota,\rho,\rho)$. 
For such $e$, we define an adjoint channel
$\bar e:=(\iota,\bar\rho,\bar\rho)$.   
Following~\cite{FRSII}, we define a pseudo-adjoint $F\mapsto F^\adj$ 
on the space $\F_\iota$ by 
\begin{equation*}
F(e,A)^\adj := F(\bar e, \bar\rho(A^*)R_\rho), 
\quad  e=(\iota,\rho,\rho).
\end{equation*}
This pseudo-adjoint does {\em not} coincide with the operator adjoint (with
respect to which $\F$ is a $C^*$-algebra). In fact, it is not an
involution, but rather satisfies 
\begin{equation} \label{eqF**}
(F^\adj)^\adj = \chi_\rho\, F 
\end{equation}
if $F$ has charge $\rho$. 
The number $\chi_\rho$ is a root of unity in a self-conjugate sector 
($\barrho\simeq\rho$), intrinsic to the sector, while in all other
sectors $R_\rho$ and $R_\barrho$ may be chosen so that
$\chi_\rho=1$~\cite[Eq.~(3.2)]{FRSII}.  
(Those self-conjugate sectors with $\chi_\rho=-1$ are called 
pseudo-real sectors.) 
The adjoint preserves localization, i.e.\ leaves invariant the spaces
of local fields with trivial source, $\F_\iota(\cccpath):=
\F_\iota\cap\F(\cccpath)$: 
\begin{equation*} 
\big(\F_\iota(\cccpath)\big)^\adj = \F_\iota(\cccpath).
\end{equation*}
Finally, the adjoint is preserved by Poincar\'e transformations: 
\begin{equation*} 
\big(U(\potild) F U(\potild)^*\big)^\adj=
U(\potild) F^\adj U(\potild)^*
\end{equation*}
for all $\potild\in\Potild$ and $F\in\F_\iota$. 
Due to the faithfulness of $\pi_0\rho$ on the local algebras $\Auni(\ccc)$, 
$F^\adj\Omega=0$ implies $F=0$ for $F\in\F(\cccpath)$. This allows for
the definition of our pseudo-Tomita operator~\cite{FRSII}  
\begin{equation}  \label{eqSTom} 
\SF:\, F\Omega \mapsto F^\adj\Omega, \quad F\in\F_\iota(\We). 
\end{equation} 
Here, $\We$ is a path ending at $W_1$ which will be specified in 
Eq.~\eqref{eqWe} below. 
\paragraph{{\it Braid Group Statistics.}} 
For every pair of localized morphisms $\rho,\sigma$ in $\Sec$ there is
a local unitary intertwiner $\eps(\rho,\sigma)\in
\Int(\sigma\rho|\rho\sigma)$, the so-called statistics operator. The
family of statistics operators satisfies the braid
relations~\cite[Eq.~(2.6)]{FRSII} 
and determines the statistics of fields, as follows. 
Let $\spc_1$ and  $\spc_2$ be causally separated, and let
$\spcpath_i$ be paths ending at $\spc_i$ with relative winding number 
$N(\spcpath_2,\spcpath_1)=n$, and let $F(e_1,A_1)\in\F(\spcpath_1)$ and 
$F(e_2,A_2)\in\F(\spcpath_2)$ be two fields with superselection
channels $e_1$ of type $(\alpha,\rho_1,\beta)$ and $e_2$ of type
$(\beta,\rho_2,\gamma)$, where $\alpha,\beta,\gamma,\rho_i$
$\in\Sec$. Then there holds the commutation
relation\cite[Prop.~5.9]{FRSII} 
\begin{equation} \label{eqFieldCR}
F(e_2,A_2)\, F(e_1,A_1) = \sum_{\delta,i_1,i_2}  
R(\delta,e_1,e_2,n) 
\,F(\hat e_1,A_1)\, F(\hat e_2,A_2).   
\end{equation}
Here $\hat e_1=(\delta,\rho_1,\gamma,i_1)$ and $\hat
e_2=(\alpha,\rho_2,\delta,i_2)$, and the sum goes over all 
morphisms $\delta$ which are contained in the product
representation $\alpha\circ \rho_2$. The numbers $R(\cdot)$ are given
by
\begin{equation} \label{eqRMat}
R(\delta,e_1,e_2,n)= 
\big(\frac{\omega_\alpha\omega_\gamma}{\omega_\beta\omega_\delta}\big)^{n}\;
\pi_0\big(T_{e_2}^*T_{e_1}^*\alpha(\eps(\rho_2,\rho_1))T_{\hat
e_2}T_{\hat e_1}\big). 
\end{equation}
The vacuum expectation values~\eqref{eqCommut} of these commutation 
relations are already determined by the statistics phases. 
The statistics parameter $\lambda_\rho$ and statistics phase 
$\omega_\rho$ of a sector $[\rho]$ are defined by the relations  
\begin{equation} 
\phi_\rho(\eps(\rho,\rho))=\lambda_\rho \unity, \qquad \omega_\rho =
\frac{\lambda_\rho}{|\lambda_\rho|},  
\end{equation}
respectively. (They depend only on the equivalence class of $\rho$.)
Suppose now that $\spc_1$ and $\spc_2$ are
causally separated, and $\spcpath_1$ and $\spcpath_2$ have relative winding
number $N(\spcpath_2,\spcpath_1)=-1$, see Figure~2 for an example. 
Then for $F_1=F(e,A_1)$ and $F_2=F(e,A_2)\in\F_\iota(\spcpath_i)$  with
$e=(\iota,\rho,\rho)$ there holds 
\begin{equation}\label{eqCommut}
\big(\, F_2\Omega,F_1\Omega\,\big) = 
\omega_\rho \,\big(\, F_1^\adj\Omega,F_2^\adj\Omega\,\big), 
\end{equation}
see e.g.~\cite[Eq.~(6.5)]{DHRIV} and \cite[Lemma~A.1]{Mu_SpiSta}.  
Of course $\omega_\rho=\pm1$ corresponds to Bosons/ Fermi\-ons, while the
generic case 
corresponds to braid group statistics. 
Note that the hypothesis on the relative winding number under which 
Eq.~\eqref{eqCommut} holds is not symmetric in $\spcpath_1$ and
$\spcpath_2$. 
Without this condition, Eq.~\eqref{eqCommut} would imply
$\omega_\rho\omega_\barrho=1$. But $\omega_\rho$ and
$\omega_\barrho$ are known to coincide~\cite{FM2}, hence 
Eq.~\eqref{eqCommut} would be be self-consistent only
for $\omega_\rho=\pm 1$, excluding braid group statistics.  
\paragraph{Assumptions.} 
We shall assume that the theory is purely massive, and that asymptotic
completeness holds. By purely massive, we mean that the set $\Sec$ of
relevant sectors is generated by a set of elementary charges, 
which correspond to massive particles. 

A covariant representation  is called a 
{\em massive single particle representation} if its mass 
spectrum\footnote{By mass spectrum we mean the spectrum of the mass
operator $P_\mu P^\mu$.} 
contains a strictly positive eigenvalue (the mass of the corresponding
particle type), isolated from the rest of the mass 
spectrum in its sector by a mass gap (implementing the
idea that there are no massless particles in the model).  
We also assume that there are only finitely many particle
types in a given sector with a given mass, and that these have the same spin. 
We thus make the 
\begin{Ass}[Massive particle spectrum.] \label{A1}
There is a finite subset 
$$
\SecE\subset \Sec
$$
of morphisms corresponding to massive single particle
representations, which generates $\Sec$.  (In other words,
$\Sec$ is exhausted by composition and subsequent reduction of
morphisms in $\SecE$.) 
For each $\rho\in\SecE$, the restriction of the representation 
$U_\rho$ to the eigenspace of the corresponding mass
value\footnote{For simplicity, we shall assume that there is only one 
mass eigenvalue in each sector, but our results still hold if no
restriction is imposed on the number of (isolated) mass values in each
sector.} $m_\rho$ is 
\label{eqFinite} 
a finite multiple of an irreducible representation. 
\end{Ass}
(Note that $\Sec$ is countable but may be infinite.) 
It is gratifying that this assumption, together with Haag 
duality~\eqref{eqHD}, implies that all relevant sectors really 
have as representatives localized morphisms~\cite{BuF} with finite 
statistics~\cite{F81} as assumed in our framework. 
Our second assumption is that the theory can be completely interpreted
in terms of multi-particle states: 
\begin{Ass}[Asymptotic Completeness.] \label{A2}
The scattering states span the entire Hilbert space $\calH$.  
\end{Ass}
(We shall sketch in Section~\ref{secHex} the Haag-Ruelle construction
of scattering states from single particle states 
in the setting of the reduced field bundle for Plektons.)  
\section{Algebraic Properties of the Pseudo-Modular Objects}  \label{secAlg} 
As a first step, we discuss algebraic properties of the pseudo-modular 
objects which are independent of our special 
assumptions \ref{A1} and \ref{A2}, and are analogous to properties of
genuine modular objects. 
In particular, we show that the (adjoint action of the) pseudo-modular
group leaves the field algebra of the wedge invariant, and 
point out that Borchers' commutation relations between the modular objects 
and the translations also hold in the present case. 
To this end, we exhibit the pseudo-Tomita operator $\SF$ as a family
of {\em relative} Tomita operators~\cite{Stratila} $S_\rho$,
$\rho\in\Sec$,  associated
with the observable algebra and certain suitably chosen pairs of states. 
We then use results obtained in a recent article~\cite{Mu_BorchersCR} 
by the author on the relative modular objects.

Recall that $\SF$ maps each $\calH_\rho$ to $\calH_\barrho$,  
and therefore corresponds to a family  of operators $S_\rho$ 
acting in $\HO$ via 
\begin{equation} \label{eqModRho} 
S(\rho,\psi) =: (\barrho,S_\rho \psi). 
\end{equation}
In order to calculate each operator $S_\rho$ explicitly, we first have to
specify the path $\We$ from $\spc_0$ to $W_1$ which enters in its 
definition~\eqref{eqSTom}.  
For simplicity, we shall take the reference cone $\spc_0$ 
to be properly contained in $W_1$ (i.e., its closure is contained in
$W_1$), and define $\We$ to be the (class of the) path 
\begin{equation} \label{eqWe}
\We := (\spc_0,W_1). 
\end{equation}
Then, $F(e,A)$ is in $\F_\iota(\We)$ if and only if $A$ is in
$\Auni(W_1)$ and  $e$ is of the form $e=(0,\rho,\rho)$ for some 
$\rho\in\Sec$, in which case $T_e=R_\rho$. Thus the
definition~\eqref{eqSTom} reads explicitly
\begin{equation*}  
\SF_\rho \, \pi_0(A)\OmO = 
\pi_0[\barrho(A^*)R_\rho]\OmO, \quad  A\in\Auni(W_1).
\end{equation*} 
In the special case $\rho=\iota$, this is just the
Tomita operator of the observables, which we shall denote by 
$\SO\equiv \JO\DO^{1/2}$. 
In the general case $\rho\neq \iota$, $\SF_\rho$ is the relative Tomita 
operator associated with the algebra $\Auni(W_1)$ and the pair of
states $\omega_0:=\lsp \OmO,\pi_0(\cdot)\OmO\rsp$ (the vacuum 
state) and the positive functional 
$$
\varphi_\rho:= |\lambda_\rho|^{-1}\; \omega_0\circ \phi_\rho 
= \big(R_\rho\OmO,\pi_0\bar\rho(\cdot) R_\rho\OmO\big).  
$$
We denote again the polar decomposition  by
$S_\rho=J_\rho\Delta_\rho^{1/2}$ and call $\Delta_\rho^{it}$ and
$J_\rho$ the relative modular group and conjugation, 
respectively. 
It is known in relative Tomita-Takesaki theory~\cite{Stratila} that 
the operator $\Delta_\rho^{it}\Delta_0^{-it}$ is in 
$\pi_0\Auni(W_1)$ for $t\in\RR$, giving rise to a family of unitaries 
\begin{equation} \label{eqCoc}
\coc_\rho(t):= \pi_0^{-1}\big(\Delta_\rho^{it}\Delta_0^{-it}\big) 
\quad \in \Auni(W_1)  
\end{equation} 
known as the Connes cocycle $(D\varphi_\rho:D\omega_0)_t$ with respect
to the pair of weights $\omega_0$ and $\varphi_\rho$. 
Starting from the Connes cocycle, the algebraic properties of the 
relative modular objects have been analyzed in~\cite{Mu_BorchersCR}. 
It has been shown there that the relative modular unitary group 
$\Delta_\rho^{it}$ and conjugation
$J_\rho$ satisfy the implementation properties 
\begin{align}  \label{eqDeltaRhoSig} 
\Ad \Delta_\rho^{it}\circ\pi_0\rho&=\pi_0\rho\circ\sigma_t, \\
\Ad J_\rho\circ \pi_0\rho &= \pi_0\barrho\circ \alphaO_j \nonumber
\end{align}
on $\Auni(W_1)\cup\Auni(W_1')$. 
Here, $\sigma_t$ is the modular group associated with  
$\Auni(W_1)$ and the vacuum state~\cite{BraRob}, and  $\alphaO_j$ is
an anti-isomorphism from $\Auni(W_1)$ onto $\Auni(W_1')$ and vice
versa. These are characterized by the fact that there holds
\begin{align}  
\Ad \DO^{it}\circ\pi_0&=\pi_0\circ\sigma_t \nonumber\\
\Ad J_0\circ \pi_0 &= \pi_0\circ \alphaO_j\label{eqJ0alphaj} 
\end{align}
on $\Auni(W_1)\cup\Auni(W_1')$. 
Furthermore, Borchers' commutation relations  have been 
shown~\cite{Mu_BorchersCR} to hold between the relative modular objects
and the translations, namely for all $t\in\RR$ and $x\in\RR^3$ there holds 
\begin{align} 
\Delta_\rho^{it}\,U_\rho(x,\unit)\,\Delta_\rho^{-it}&= 
U_\rho(\Boox{-2\pi t}x,\unit), 
\label{eqDUD}\\
J_\rho\,U_\rho(x,\unit)\,J_\rho^{-1} &= U_\barrho(jx,\unit). \label{eqJUJ} 
\end{align}
Finally, the algebraic relations among the relative modular objects,  
also established in~\cite{Mu_BorchersCR}, are completely analogous to the 
case of the genuine ones: 
\begin{align}
J_\rho\;\Delta_\rho^{it}\; J_\rho^{-1}&=\Delta_\barrho^{it},
\label{eqJDJ} \\
J_\rho\,J_\barrho &=\chi_\rho \unity, \label{eqJJ} 
\end{align}
where $\chi_\rho$ are the factors defined after Eq.~\eqref{eqF**}. 
Note that these equations (or Eq.~\eqref{eqF**}) 
imply that the pseudo-Tomita operator $\SF$ 
of the field algebra is not an involution, but rather satisfies 
\begin{equation} \label{eqS2}
\SF^2  \subset \chi \; := \, \sum_{\rho\in\Sec} \chi_\rho\,E_\rho.  
\end{equation}

Using relative Tomita-Takesaki theory and the fact that the
pseudo-modular objects are related to the relative ones by 
\begin{equation} \label{eqModRho'} 
\JF(\rho,\psi)= (\barrho,J_\rho\psi), \quad 
\DF^{it}(\rho,\psi)= (\rho,\Delta_\rho^{it}\psi) 
\end{equation}
due to uniqueness of the polar decomposition, we shall now calculate the
adjoint action of the pseudo-modular group on the fields localized in 
$\We$. It turns out
that this action leaves $\F(\We)$ invariant --- a non-trivial fact which
enters crucially in the calculation of $\DF^{it}$ on 
scattering states in Section~\ref{secHex}. For completeness' sake we
also show that the action of the pseudo-modular group commutes with 
the pseudo-adjoint $\adj$. 
\begin{Prop} \label{FAInv}
The adjoint action of the pseudo-modular group leaves the field 
algebra $\F(\We)$ associated to the wedge invariant, and commutes with
the pseudo-adjoint~$\adj$ on $\F_\iota(\We)$:  
\begin{align} \label{eqDeltaW1}
 \DF^{it}\;\F(\We)\; \DF^{-it} & = \F(\We),\\
\label{eqModAdj}
  (\DF^{it}F\DF^{-it})^{\adj} & = \DF^{it} F^{\adj} \DF^{-it},
\end{align}
$F\in\F_\iota(\We)$.  
Specifically, if $e$ is a superselection channel with $c(e)=\rho$, then 
\begin{equation} \label{eqModField}
 \DF^{it}\,F(e,A)\, \DF^{-it} = F(e,\coc_\rho(t)\,\sigma_t^0(A)).
\end{equation} 
\end{Prop}
It is interesting to note the resemblance of Eq.s~\eqref{eqCoc} and
\eqref{eqModField} with Eq.s ~\eqref{eqYrho} and \eqref{eqUImplement},
respectively. 
\begin{Proof}
We shall use two facts about the Connes cocycle~\eqref{eqCoc} 
established by Longo 
in the present context. Namely, on 
$\calA(W_1)\cup \Auni(W_1')$ there holds~\cite[Prop.\ 1.1]{Longo97}
\begin{equation} \label{eqCocInt} 
\Ad \coc_\rho(t)\circ \sigma_t \circ \rho  = \rho \circ \sigma_t, 
\end{equation}
and for any intertwiner $T^*$ from $\rho_r$ to $\rho_s\rho_c$
there holds~\cite[Props.\ 1.3, 1.4]{Longo97}  
\begin{equation} \label{eqLongo2}
T^* \;\rho_s(\coc_{\rho_c}(t))\,\coc_{\rho_s}(t)= 
\coc_{\rho_r}(t)\,\sigma_t^0(T^*). 
\end{equation}
For the proof of the proposition, let $A$ be in $\Auni(W_1)$ and $e$
be of type $(\rho_s,\rho_c,\rho_r)$. There holds 
\begin{align*}
 \DF^{it}\,F(e,A)\, \DF^{-it}\, (\rho_s,\psi) &= 
\big(\rho_r,\Delta_{\rho_r}^{it}
\pi_0(T_e^*\rho_s(A))\Delta_{\rho_s}^{-it}\psi\big) \\
&= \big(\rho_r,\pi_0(T_e^*\rho_s(\coc_{\rho_c}))\Delta_{\rho_s}^{it}
\pi_0\rho_s(A)\Delta_{\rho_s}^{-it}\psi\big)\\
&=\big(\rho_r,\pi_0(T_e^*\rho_s(\coc_{\rho_c}))
\pi_0\rho_s(\sigma_t^0(A))\psi\big)\\
&= F(e,\coc_{\rho_c}(t)\,\sigma_t^0(A))\, (\rho_s,\psi).
\end{align*}
In the third equation we have used Eq.~\eqref{eqDeltaRhoSig}, and in
the second one we have used that 
\begin{equation*} 
\Delta_{\rho_r}^{it}\pi_0(T^*) = 
\pi_0(T^* \rho_s(\coc_{\rho_c}(t))\, \Delta_{\rho_s}^{it}, 
\end{equation*}
which is a consequence of Eq.~\eqref{eqLongo2}. 
This proves the explicit formula~\eqref{eqModField}. 
Since $\coc_{\rho_c}(t)$ and $\sigma_t(A)$ both are in $\Auni(W_1)$,
this also shows invariance~\eqref{eqDeltaW1} of $\F(\We)$. 
To prove Eq.~\eqref{eqModAdj}, let $e=(\iota,\rho,\rho)$ and 
$\bar{e}:=(\iota,\barrho,\barrho)$. Then 
\begin{align*}
 (\DF^{it}\,F(e,A)\, \DF^{-it})^\adj &= 
(\bar e,\barrho[\sigma_t^0(A^*)\coc_\rho(t)^*]R_\rho),\\
 \DF^{it}\,F(e,A)^\adj\, \DF^{-it} &= 
(\bar e,\coc_\barrho(t)\sigma_t^0[\barrho(A^*)R_\rho]). 
\end{align*}
But Eq.s~\eqref{eqCocInt} and \eqref{eqLongo2} imply that 
\begin{align*}
\barrho[\sigma_t^0(A)\coc_\rho(t)^*]R_\rho &=
\coc_\barrho(t)\sigma_t^0\barrho(A)\coc_\barrho(t)^*
\barrho(\coc_{\rho}(t)^*)R_\rho = 
\coc_\barrho(t) \sigma_t^0[\barrho(A)R_\rho].
\end{align*}
This shows Eq.~\eqref{eqModAdj} and completes the proof. 
\end{Proof}
\section{Modular Covariance and \CPT operator 
on the Single Particle Space} 
 \label{secH1}
As a first step, we prove single-particle versions of the 
Bisognano-Wichmann and the CPT theorems.  For $\rho$ in the set
$\SecE$ of single particle charges, let $E_\rho$ be the
projection from $\calH$ onto $\calH_\rho$, and $\Ee_\rho$ the
projection from $\calH$ onto the eigenspaces of the mass operator in
$\calH_\rho$ (corresponding to the isolated eigenvalues in the sector
$\rho$). 
We denote by $\Ee$ the sum of all $\Ee_\rho,$ where 
$\rho$ runs through $\SecE$, and call the range of $\Ee$ the single 
particle space. 

Borchers' commutation relations~\eqref{eqDUD}, \eqref{eqJUJ} imply that 
the pseudo-modular unitary group and the pseudo-modular conjugation 
commute with the mass operator. Hence the pseudo-Tomita operator $\SF$ 
commutes with $\Ee.$ Let us denote the corresponding restriction by 
\begin{equation*} 
 \SFE:= \SF\,\Ee \,. 
\end{equation*} 
Similarly, the representation $U(\Potild)$  leaves $\Ee\calH$ invariant,
giving rise to the sub-representation 
\begin{equation*} 
  \Ue(g):= U(g)\,\Ee\,, 
\end{equation*}    
and one may ask if modular covariance holds on $\Ee\calH.$ We
show in this section that this is indeed the case, the line of
argument being as follows. Let $K$ denote the generator of the unitary
group of 1-boosts, $\Ue(\lambda_1(t))=\exp(itK).$ We exhibit in
Eq.~\eqref{eqUj1} below an anti-unitary 
``\CPT-operator'' $\Uej$ representing the reflexion $j$ on $\Ee\calH,$ and 
show that  $\SFE$ coincides with the ``geometric'' involution 
\begin{equation} \label{eqSgeo}
\SgeoE:= \Uej\;e^{-\pi K}\, 
\end{equation} 
up to a unitary operator which commutes with the representation 
$\Ue$ of $\Potild$. By uniqueness of the polar decomposition, this
will imply modular covariance on $\Ee\calH$, namely 
$\Ue(\Boox{-2\pi t})\equiv \exp(-2\pi it K)=\Delta^{it} \Ee$.  

We begin by exploiting our knowledge about $\Ue(\Potild).$ By assumption, for
each $\rho\in\SecE$ the sub-representation $\Ue_\rho:=E_\rho \Ue$  
is equivalent to a finite number, say $n_\rho$, of copies of
the irreducible ``Wigner'' representation of the universal covering of the 
Poincar\'e group with mass $m_\rho$ and a certain spin $s_\rho\in\RR$. 
Let us denote this representation by $\UWig_\rho$. 
It acts on the Hilbert space $L^2(H_{m_{\rho}}^+,d\mu)\otimes \CC^{n_\rho}$,
which consists of momentum space ``wave functions''  $\psi:H_{m_\rho}^+ \to
\CC^{n_\rho}$ living on the mass shell 
$H_{m_\rho}^+:= \{ p\in \RR^3|\,p\cdot p = m_\rho^2, \,p_0>0\}$  
and having finite norm w.r.t.\ the scalar product 
$$
\big(\psi,\phi \big) =\int_{H_{m_\rho}^+}d\mu(p) 
\big(\psi(p)\,,\,\phi(p)\big)_{\CC^{n_\rho}}. 
$$
The representation $\UWig_\rho$ acts in this space as (see
e.g.~\cite{M02a})  
\begin{equation*} 
\big(\UWig_\rho(a,\lortild)\psi\big) (p) = 
e^{is\WigRot(\lortild,p)}\,e^{ia\cdot p}\,
\psi(\lor^{-1}p)\,, 
\end{equation*}
where $\lor$ is the Lorentz
transformation onto which $\lortild$ projects, and 
$\WigRot(\lortild,p)\in\RR$ is the so-called Wigner rotation. 
To the representation $\UWig_\rho$ an anti-unitary operator
$\UWig_\rho(j)$ can be adjoined satisfying the representation properties 
\begin{equation} \label{eqUWjgj1}
  \UWig_\rho(j)^2=\unity\;\quad\text{ and }\quad 
  \UWig_\rho(j)\,\UWig_\rho(\potild)\,\UWig_\rho(j)=\UWig_\rho(j\potild j) 
\end{equation}
for all  $\potild\in\Potild.$ Namely, it is given 
by~\cite{M02a}
\begin{equation*} 
\big(\UWig_\rho(j)\psi\big)(p):= \overline{\psi({-j}\, p)}\,, 
\end{equation*} 
where the overline denotes component-wise complex conjugation in
$\CC^{n_\rho}$. $\UWig_\rho$ is then a representation of the group
$\Potildj$ which we identify with the semi-direct product of 
$\Potild$ and $\II_2$,
the latter acting in the former via the unique lift~\cite{Var2} of the 
adjoint action of $j$ on $\Po$. 
Let us denote the unitary intertwiner between the representations 
$\UWig_\rho$ and $\Ue_\rho$ of $\Potild$ by $W_\rho$. In other words, $W_\rho$
is an isometric isomorphism from $L^2(H_{m_\rho}^+,d\mu)\otimes
\CC^{n_\rho}$  onto 
$\Ee_\rho\calH$ satisfying 
\begin{equation}  \label{eqIntW}
\Ue_\rho(\potild)\,W_\rho =W_\rho \, \UWig_\rho(\potild) 
\end{equation}
for all $\potild\in\Potild$. 
It is known that the masses\cite{F81}, spins~\cite{Mu_SpiSta} and 
degeneracies $n_\rho$\cite{Mu_SpiSta} coincide for $\rho$ and 
$\barrho$. Hence the ranges of $W_\rho^*$ and $W_\barrho^*$ coincide,
and $C_\rho:=W_\barrho W_\rho^*$ is a unitary operator from 
$\Ee_\rho\calH$ onto $\Ee_\barrho\calH$, representing ``charge 
conjugation'', which intertwines the representations $\Ue_\rho$ and
$\Ue_\barrho$. 
Therefore the ``\CPT''-operator 
\begin{equation} \label{eqUj1}
\Ue(j):= \sum_{\rho\in\SecE} \Ue_\rho(j), \quad 
\Ue_\rho(j):= W_\barrho\UWig_\rho(j)W_\rho^*\;\equiv 
C_\rho \, W_\rho\UWig_\rho(j)W_\rho^*, 
\end{equation}
not only conjugates the charge, but also represents the reflection
$j$, namely satisfies 
\begin{equation} \label{eqUjgj1}
\Ue(j)^2 = \unity,\quad \Ue(j)\,\Ue(\potild)\,\Ue(j)^*=\Ue(j\potild j).
\end{equation}
We define now a closed anti-linear operator $\SgeoE$ in terms of the
representation $\Ue(\Potildj)$, as anticipated, by
Eq.~\eqref{eqSgeo}. 
Note that the group relation $j\,\boox{t} \,j=\boox{t}$ implies that
$\SgeoE$ is an involution: it leaves its domain invariant and
satisfies $(\SgeoE)^2\subset\unity$. 
\begin{Prop}  \label{CovH1} 
There is a unitary operator $D$ on $\Ee\calH$ commuting with the
representation $\Ue$ of $\Potild$ and with each $E_\rho$, such that  
\begin{equation} \label{eqCSS}
   \SgeoE =D \, \SFE\,.
\end{equation}
\end{Prop}
\begin{Proof}
Let $\Spc_1$ denote the class of space-like cones which have apex at the
origin, contain the positive $x^1$-axis and are contained in the wedge
$W_1$, and have non-zero intersection with the
time-zero hyper-surface, and let $\Spcpaths_1$ be the set of
(equivalence classes of) paths in $\Ccc$ of the form $(\spc_0,W_1,\spc)$
with $\spc\in\Spc_1$. Let further
$\F_{\iota,\rho}^\infty(\spcpath)$ be the set of field operators 
$F\in\F_\iota(\spcpath)$ with superselection channel
$e=(\iota,\rho,\rho)$ and for which $\potild\mapsto \alphaO_{\potild}(F)$
are smooth functions. 
\begin{Lem}[\cite{BuEp,Mu_SpiSta}]
Let $\spcpath\in\Spcpaths_1$ and $F\in\F_{\iota,\rho}^\infty(\spcpath)$. 
Then, for fixed $p\in H_{m_\rho}^+$, the $\CC^{n_\rho}$-valued
function 
\begin{equation} \label{eqPsit}
t\mapsto \psi(t,p):= \big(\UWig_\rho(\boox{t}) W_\rho^*\Ee_\rho
  F \Omega\big)(p)
\end{equation}
extends to an analytic function in the strip $t\in \RR+i(0,\pi)$,
which is continuous and bounded on its closure. 
At $t=i\pi$, it has the boundary value 
\begin{equation} \label{eqPsiipi}
\overline{\psi(t,-jp)|_{t=i\pi}} =
D_\barrho\,\big(W_\barrho^*\Ee_\barrho F^\adj\Omega\big)(p),   
\end{equation}
where $D_\barrho$ is an isometry in $\CC^{n_\rho}$ independent of
$\spcpath$ and $F$.\footnote{On the left hand side of
 Eq.~\eqref{eqPsiipi}, one first analytically continues into $t=i\pi$ and
 then conjugates component-wise in $\CC^{n_\rho}$.}
\end{Lem}
\begin{Proof} 
We show how this lemma follows from~\cite[Proposition~2]{Mu_SpiSta}, 
which in turn is based on the work of Buchholz and Epstein~\cite{BuEp}. 
If $\spcpath\in\Spcpaths_1$, then $\spc$ contains a space-like
cone $\spc_1$ of the special class used in \cite{BuEp} and
\cite[see Eq.~(16)]{Mu_SpiSta}. 
Defining $\spc_2:=-\spc_1$ and choosing a path $\spcpath_2$ 
ending at $\spc_2$ and satisfying $N(\spcpath_2,\spcpath_1)=-1$, 
all conditions on $\spcpath_1$
and $\spcpath_2$ used in~\cite{Mu_SpiSta} are satisfied. (In 
particular, the ``dual''~\cite[Eq.~(17)]{Mu_SpiSta} of the 
difference cone $\spc_2-\spc_1$ contains the negative $x^1$-axis, as
required in Eq.~(36) of \cite{Mu_SpiSta}.) 
For $i\in\{1,2\}$ we now pick $n_\rho$ operators
$F_{i,\beta}\in\F_{\iota,\rho}^\infty(\spcpath_i)$, $\beta=1,\ldots,n_\rho$, 
with $F_{1,1}:=F$ of the lemma, such that the $n_\rho$ vectors 
$$ 
\big(W_\rho^*\Ee_\rho F_{i,\beta}\Omega\big)(p)\;\in\CC^{n_\rho} 
$$
are linearly independent for all $p$ in
some open set.\footnote{If $\Ee_\rho F\Omega$ is non-zero, then this
  is possible due to the Reeh-Schlieder property.}
Then Proposition~2 in \cite{Mu_SpiSta}  asserts 
that there is an isometric  matrix $D_\barrho$, independent of 
$\spcpath_i$ and $F_{i,\beta}$, such that 
the assertion of our Lemma holds for all $F_{1,\beta}$. 
This completes the proof of the lemma. 
\end{Proof}
We now reformulate the lemma in terms of the operators $\SFE$ and $\SgeoE$. 
By the lemma,  there is an operator 
$A_\rho$ on $\Ee_\rho\calH$ with domain 
\begin{equation*}
\calD_{0,\rho}^{(1)}:=\underset{\spcpath\in\Spcpaths_1}{{\rm span }} \, 
\Ee\,\F_{\iota,\rho}^\infty(\spcpath)\,\Omega   
\end{equation*}
defined via 
\begin{equation} 
\big(W_\rho^* A_\rho \phi\big)(p) :=
\big(W^*_\rho\Ue_\rho(\boox{t}) \phi\big)(p)\big|_{t=i\pi}, 
\quad \phi \in \calD_{0,\rho}^{(1)}. 
\end{equation}
Denoting by $\hat D_\rho$ the multiplication operator with the 
matrix $D_\rho$, 
$$
\big(W_\rho^*\,\hat D_\rho\phi\big)(p) :=
D_\rho\,\big(W_\rho^*\phi\big)(p), 
$$
Eq.~\eqref{eqPsiipi} of the Lemma reads 
\begin{equation} \label{eqJADS}
\Ue_\rho(j)\,A_\rho \subset \hat D_\barrho \SFE \,E_\rho. 
\end{equation}
(Note that the domain $\calD_{0,\rho}^{(1)}$ of $A_\rho$ is contained
in the domain of $\SFE$.)  
We wish to identify $A_\rho$ with $e^{-\pi K_\rho}$, where $K_\rho$ is
the generator of the one-parameter group $\Ue_\rho(\boox{t})$. First,
relation~\eqref{eqJADS} shows that $A_\rho$ is closable since $\SFE$ is. 
Now for $\phi$ in the dense domain of $A_\rho^*$, 
$\psi\in \calD_{0,\rho}^{(1)}$ and $f\in C_0^\infty(\RR)$ one finds
(by a calculation as in~\cite[proof of Lemma~11]{M01a}): 
$$
\big(\phi\,, e^{-\pi K_\rho} f(K_\rho) \psi\big) = 
\big(A_\rho^* \phi\,, f(K_\rho) \psi\big). 
$$
This implies that the span of vectors of the form $f(K_\rho)\psi$ as above, 
$\calD_{\rho}^{(1)}$, is in the domain of the closure $A_\rho^{**}$
of $A_\rho$, and that on $\calD_{\rho}^{(1)}$ the closure $A_\rho^{**}$
coincides with  $e^{-\pi K_\rho}$.   
But $\calD_{\rho}^{(1)}$ is invariant under the unitary one-parameter
group $\Ue_\rho(\boox{t})$, because for each $t$ there is some 
$\spcpath_t\in\Spcpaths_1$ such that $\Boox{t}\spcpath\subset \spcpath_t$. 
By standard arguments, the closure of $A_\rho$ then coincides with the 
operator $e^{-\pi K_\rho}$. 
In particular, $\calD_{0,\rho}^{(1)}$ is a core for $e^{-\pi K_\rho}$,
$\rho\in\SecE$, hence $\bigoplus_\rho \calD_{0,\rho}^{(1)}$ is a core for
$\SgeoE$. Therefore, relation~\eqref{eqJADS} implies 
\begin{equation} \label{eqSgeoinS}
\SgeoE \subset  D\,\SFE, 
\end{equation} 
where $D:=\bigoplus_\rho \hat{D}_\rho$. 
By construction, $D$ commutes with $E_\rho$ and with the
representation $\Ue$, as claimed in the Proposition. 
It remains to show the opposite inclusion ``$\supset$'' in
Eq.~\eqref{eqSgeoinS}. 
To this end, we refer to
the opposite wedge $W_1'=\Rot{\pi}\, W_1$. Let 
\begin{align*}
\SFL&:=U(\rot{-\pi})\;\SF  \;U(\rot{\pi})\,, \\
\SFEL&:=\Ue(\rot{-\pi})\;\SFE  \;\Ue(\rot{\pi}) \equiv \SFL\,\Ee\,,\\
\SgeoEL&:=\Ue(\rot{-\pi})\;\SgeoE\;\Ue(\rot{\pi})\,.
\end{align*}
We claim that the following sequence of relations holds true: 
\begin{equation} \label{eqRLLR}
\SFE \subset \omega\, \chi \,(\SFEL)^*
\subset \omega \,D^* \,(\SgeoEL)^*=\omega \, 
\Ue(\rot{-2\pi})D^* \,\SgeoE\;.
\end{equation}
Here, $\omega:=\sum_\rho \omega_\rho E_\rho$ where 
$\omega_\rho$ are the statistics phases, and 
$\chi$ is the operator defined in Eq.~\eqref{eqS2}. 
To see the first inclusion, note that $\SFL$ is the closure of the
operator $F\Omega\mapsto F^\adj\Omega$,
$F\in\F_\iota(\rot{-\pi}\We)$. Therefore Eq.s~\eqref{eqCommut} and 
\eqref{eqF**} imply that for $F_1\in\F_\iota(\We)$ and
$F_2\in\F_\iota(\rot{-\pi}\We)$, there holds 
$$
\big(\,\omega \,\chi^*\, \SFL\, F_2^\adj\Omega,F_1\Omega\,\big) = 
\big(\, \SF \, F_1\Omega,F_2^\adj\Omega\,\big).
$$ 
This implies $\SF \subset (\omega \chi^*\SFL)^*\equiv \SFL^*\chi\omega^*$, 
which coincides with $\omega \chi \SFL^*$ because 
$\chi_\barrho=\overline{\chi_\rho}$ and $\omega_\barrho=\omega_\rho$,
while  $\SFL^*$ is anti-linear and maps $E_\rho\calH$ into
$E_\barrho\calH$. Since $\SF$ commutes with $\Ee$ due to Borchers'
commutation relations, this implies the first inclusion in \eqref{eqRLLR}.   
To show the second inclusion, we first note that
relations~\eqref{eqSgeoinS} and~\eqref{eqS2} imply
$(D^*\SgeoE)^2\subset(\SFE)^2\subset  \chi$. 
Since $\SgeoEL=(\SgeoEL)^{-1}$, this yields  
\begin{equation*} 
D^*\,\SgeoE = \chi\, \SgeoE\,D \equiv \SgeoE\,D\,\chi 
\end{equation*}
(we have equality here instead of $\subset$, since $D$ leaves the
domain of $\SgeoE$ invariant), and the same relation holds for $\SgeoEL$. 
Therefore, the adjoint of Eq.~\eqref{eqSgeoinS} yields 
\begin{equation*} 
(\SFEL)^* \subset (D^*\,\SgeoEL)^* \equiv \chi^* D^* (\SgeoEL)^*, 
\end{equation*}
which implies the second inclusion in \eqref{eqRLLR}. 
As to the last equality in Eq.~\eqref{eqRLLR}, note that the group
relations $j\rot{\omega}j=\rot{-\omega}$,
$\rot{\pi}\boox{t}\rot{-\pi}=\boox{-t}$ and $j\boox{t}j=\boox{t}$ imply that 
$(\SgeoEL)^*$ 
coincides with $\Ue(\rot{-2\pi})\SgeoE$. This completes the proof of 
the sequence of relations~\eqref{eqRLLR}. Using 
relation~\eqref{eqSgeoinS} then yields  
$$
\SFE \subset \omega \, \Ue(\rot{-2\pi})\, D^* \, \SgeoE \subset 
\omega\, \Ue(\rot{-2\pi})\,\SFE. 
$$
This implies firstly that $\omega=\Ue(\rot{2\pi})$ (which is the
spin-statistics theorem) and secondly that  $\SFE\subset D^*
\SgeoE$, and completes the proof of the Proposition. 
\end{Proof} 
Note that the proof shows that, although we use results 
from~\cite{Mu_SpiSta}, the spin-statistics connection needs not be 
assumed but rather follows. 

By uniqueness of the polar decomposition, equation~\eqref{eqCSS} of the 
proposition implies the equations 
\begin{equation*} 
 \DF^{\half}\;\Ee= e^{-\pi K}\;\Ee\,,
\quad D\,\JF\;\Ee  = \Ue(j)\;\Ee \,.
\end{equation*}
Since the unitary $D$ commutes with $\Ue(\Potild)$, the above
equations and Eq.~\eqref{eqUjgj1} imply the   
single particle versions of the Bisognano-Wichmann and \CPT theorems: 
\begin{Cor} \label{CorModCov1} 
Let the Assumption~{\rm \ref{A1}} of Section~\ref{secAss} hold. Then 

\noindent i)  Modular Covariance holds on the single particle space: 
\begin{align}  \label{eqModCov1}
   \DF^{it}\;\Ee&= U(\lambda_1(-2\pi t))\;\Ee\,. \\
\intertext{ii) $\JF\;\Ee$ is a ``CPT operator'' on $\Ee\calH$, namely,
  for all $\potild\in\Potild$ holds} 
   \JF\,U(\potild)\JF^{-1} \;\Ee &=U(j\potild j) \;\Ee. 
  \label{eqJgJ1}
\end{align}
\end{Cor}
Note that $\JF$ is not an involution, but rather satisfies
$\JF^2=\chi$, c.f.~Eq.~\eqref{eqJJ}. 
\section{Modular Covariance on the Space of Scattering States} \label{secHex} 
We shall now show that modular covariance, which we have established
on the single particle
space, extends to multi-particle states via (Haag-Ruelle) scattering
theory. 
Haag-Ruelle scattering theory, as developed in~\cite{Hepp, Jost}, 
associates a multi-particle state to $n$ 
single particle vectors which are created from the vacuum by quasi-local field
operators. It has been adapted, within the field bundle formulation, 
to the setting of algebraic quantum field theory in~\cite{DHRIII}, to theories
with topological charges (i.e., charges localized in space-like cones)
in~\cite{BuF}, and to theories with braid group statistics
in~\cite{FGR}. Since we are not aware of an exposition of scattering
theory within the {\em reduced} field bundle framework, we shall give a brief 
such exposition here.  
For $\rho\in\SecE$ and $\spcpath\in\Spcpaths$, let $F=F(e,A)$ be a field
operator in $\F(\spcpath)$ carrying charge $\rho$, 
which produces from the vacuum a single particle vector with non-zero
probability in the sense that it satisfies $\Ee\,F_\iota \,\Omega \neq
0$. Here we have written 
$$
\begin{array}{lcl} 
F_\iota := F(e_\iota,A\big) &\text{ if }\; &F=F(e,A), \\   
e_\iota  := (\iota,\rho,\rho)&\text{ if }\; &e= (\rho_s,\rho,\rho_r,i). 
\end{array}
$$
The mentioned quasi-local creation operator is constructed from $F$ as
follows. 
Let $f\in\calS(\RR^3)$ be a Schwartz function whose Fourier transform 
$\tilde{f}$ has compact support contained in the open forward light cone $V_+$ 
and intersects the energy momentum spectrum of the sector $\rho$
only in the mass shell $H_{m_\rho}^+.$ 
Recall that the latter is assumed to be isolated from the rest of the
energy momentum spectrum in the sector $E_\rho\calH.$ 
Let now 
\begin{align*} 
f_t(x) & := (2\pi)^{-2}\,\int\d^3 p \:
e^{i(p_0-\omega_\rho(\bfp)) t}\,e^{-ip\cdot x}\,\tilde{f}(p), \\ 
 F(f_t) &:=\int \d^3 x \,f_t(x)\,\alpha_x (F) \;,
\end{align*}
where $\omega_\rho(\bfp):=(\bfp^2+m_\rho^2)^{1/2}.$ 
For large $|t|,$ the operator $F(f_t)$ is essentially localized
in $\spcpath+t\,V_\rho(f)$, where $V_\rho(f)$ is the velocity support
of $f$,  
\begin{equation}  \label{eqVf}
V_\rho(f):= \{\, \big(1,\frac{\bfp}{\omega_\rho(\bfp)}\big)\;,\;
p=(p^0,\bfp)\in\supp\tilde{f}\,\}\,.   
\end{equation}
Namely, for any $\eps>0$,  $F(f_t)$ can be approximated  
by an operator $F_t^\eps\in\F(\spcpath+t\,V^\eps)$, where 
$V^\eps$ is an $\eps$--neighborhood of $V_\rho(f)$, 
in the sense that  $\|F^\eps_t-F(f_t)\|$ is of fast decrease in
$t$~\cite{BBS,Hepp}.  
Further, $F_\iota(f_t)$ creates from the vacuum a single particle vector 
\begin{equation}  \label{eqFOm}
F_\iota(f_t) \,\Omega= \tilde{f}(P)\,F_\iota\,\Omega\quad\in\,\Ee_\rho\calH\,,
\end{equation}
which is independent of $t,$ and whose velocity support is contained
in that of $f.$  (Here the velocity support of a single particle
vector is defined as in Eq.~\eqref{eqVf}, with
the spectral support of $\psi$ taking the role of $\supp \tilde{f}$.) 
To construct an outgoing scattering state from $n$  single
particle vectors, 
pick $n$ localization regions  $\spcpath_i,$ $i=1,\ldots,n$ 
and compact sets $V_i$ in
velocity space, such that for suitable open neighborhoods
$V_i^\eps\subset \RR^3$ the regions 
$\spcpath_i+t\,V_i^\eps$ are mutually space-like separated for large $t$. 
Next, choose $F_i\in\F(\spcpath_i)$ with respective superselection 
channels $e_i$ suitably chosen such that the vector $F_n\cdots F_1\Omega$ does
not vanish from ``algebraic'' reasons, i.e.\ satisfying $s(e_1)=\iota$ and 
$s(e_{i}) = r(e_{i-1})$, $i=2,\ldots,n$. 
Choose further Schwartz functions $f_i$ as above with 
$V_{\rho_i}(f_i)\subset V_i.$ Then the standard lemma of
scattering theory, in the present context,  asserts the following: 
\begin{Lem} \label{Scatt}
The limit 
\begin{equation}  \label{eqBn1Om}
\lim_{t\rightarrow\infty} F_n(f_{n,t})\cdots F_1(f_{1,t})\,\Omega =: 
\big(\psi_n\times\cdots\times\psi_1\big)^{{\rm out}}\, 
\end{equation}
exists and depends only on the single particle vectors 
$\psi_i:= (F_i)_\iota(f_{i,t})\,\Omega$, on the localization regions
$\spcpath_i$ and on the superselection channels $e_i$.\footnote{We
  omit the dependence on $\spcpath_i$ and $e_i$ in our notation.}  
The limit vector is approached faster than any inverse power of $t$,  
and depends continuously on the single particle
vectors $\psi_i$. Further, there holds 
\begin{equation} \label{eqn(n-1)}
(\psi_n\times\cdots\times\psi_1)^{{\rm out}}=\lim_{t\to\infty} 
F_n(f_{n,t})\,(\psi_{n-1}\times\cdots\times\psi_1)^{{\rm out}} \,.
\end{equation}
\end{Lem}
\begin{Proof}
The space-like commutation relations~\eqref{eqFieldCR} imply that, 
for $k\in\{2,\ldots,n\}$,  
\begin{equation} \label{eqFkF1Om}
F_k(f_{k,t})\cdots F_1(f_{1,t})\,\Omega \, \simeq  \, 
\sum_{\rho_2,\ldots,\rho_{k-1}} 
R_{k-1}\cdots R_1\; F_{k-1}(f_{k-1,t})\cdots F_1(f_{1,t})\hat
F_k(f_{k,t})\,\Omega 
\end{equation}
up to terms which are of fast decrease in $t$. Here,
$R_i=R(\rho_i,e_k^{(i)},e_i,n_k^{(i)})$, where $e_k^{(i)}$ has the same charge
as $e_k$, the same source as $e_i$, and range $\rho_i$, and
$n_{k}^{(i)}=N(\spcpath_k,\spcpath_i)$. 
The sum goes over all $\rho_i$ contained in the product $c(e_k)s(e_i)$,
$i=1,\ldots,k-1$,\footnote{$\rho_1$ does not appear in the sum because
$c(e_k)s(e_1)\equiv c(e_k)$ contains no other representation
than $\rho_1\equiv c(e_k)$. In particular there is no sum if $k=2$.} 
plus the internal indices of the $e_k^{(i)}$. Further, $\hat
F_k$ is the operator arising from $F_k$ by substituting its 
superselection channel $e_k$ for $e_k^{(1)}\equiv
(\iota,\rho_1,\rho_1)$ with $\rho_1=c(e_k)$. 
{}From here the proof goes through as in~\cite{DHRIII}: 
Differentiating $F_n(f_{n,t})\cdots F_1(f_{1,t})\,\Omega$ with respect 
to $t$ yields a sum 
of terms of the form $F_n(f_{n,t})\cdots \big(\frac{d}{dt}
F_k(f_{k,t})\big)\cdots F_1(f_{1,t})\,\Omega$. Now $\frac{d}{dt}
F_k(f_{k,t})$ is of the same form as $F_k(f_{k,t})$, hence can be
permuted to the right by Eq.~\eqref{eqFkF1Om} up to fast decreasing
terms, where it annihilates the vacuum due to Eq.~\eqref{eqFOm}. This
shows the fast convergence in~\eqref{eqBn1Om}. 
Let now $G_k$ be a field operator with the same localization
$\spcpath_k$ and superselection channel $e_k$ as $F_k$ and such that
$(G_k)_\iota(g_{k})$ creates the same single particle vector
$\psi_k$ from the vacuum as $(F_k)_\iota(f_k)$. Then Eq.~\eqref{eqFkF1Om}
still holds, with the {\em same} numbers $R_i$, when $F_k(f_{k,t})$ is
replaced by  $G_k(g_{k,t})$. This implies that the scattering
state~\eqref{eqBn1Om} only depends on the single particle states,
localization regions and superselection channels. 
The continuous dependence on the single particle vectors follows from
the local tensor product structure derived in~\cite[Thm.~3.2]{FGR}. 
Eq.~\eqref{eqn(n-1)} follows  as in~\cite{Hepp} from the facts that 
$F_{n-1}(f_{n-1,t})\cdots F_{1}(f_{t})\Omega$ converges rapidly to 
$(\psi_{n-1}\times\cdots\times\psi_1)^{{\rm out}}$, while 
$\|F_n(f_{n,t})\|$ increases at most like $|t|^3$.  
\end{Proof}
Let us denote by $\calH^{(n)}$, $n\geq 2$,  the closed span of outgoing 
$n$-particle scattering states as in the lemma,  and by
$\calH^{(\Out)}$ the span of all (outgoing) particle states: 
\begin{equation*} 
\calH^{(\Out)} := \bigoplus_{n\in\NN_0} \calH^{(n)}\,.
\end{equation*}
Here $\calH^{(0)}$ is understood to be the span 
of the vacuum vector $\Omega$ and $\calH^{(1)}:=\Ee\calH$.  
{\em Asymptotic completeness} (our Assumption~{\rm \ref{A2}}) 
means that $\calH^{(\Out)}$ coincides with $\calH.$ 
Our main result is now 
\begin{Thm}[Covariance of the Pseudo-Modular Groups.] \label{ModCov}
Let the Assumptions~{\rm \ref{A1}} and {\rm \ref{A2}} of
Section~\ref{secAss} hold. 
Then the field algebra satisfies {covariance of the pseudo-modular groups}, 
\begin{equation} \label{eqModCov}
  \Delta^{it} = U(\lambda_1(-2\pi t))\,. 
\end{equation}
\end{Thm} 
(Of course, this is equivalent to 
$\Delta_\rho^{it} = U_\rho(\lambda_1(-2\pi t))$, $\rho\in\Sec$.) 
\begin{Proof} 
On the single particle space, the claim is our Corollary~\ref{CorModCov1}. 
For the scattering states, one proves Eq.~\eqref{eqModCov} by induction
over the particle number. Here, we can literally take over the proof 
from~\cite[Prop.~7]{M01a} due to the last lemma and
the results of Section~\ref{secAlg}, in
particular Borchers' commutation relations~\eqref{eqDUD} and
invariance of the wedge field algebra under the pseudo-modular
group~\eqref{eqDeltaW1}. 
Asymptotic completeness then implies that Eq.~\eqref{eqModCov} holds
on the entire Hilbert space $\calH$. 
\end{Proof}

Since for observables the pseudo-adjoint $F^\adj$ coincides with the operator
adjoint, this theorem asserts in particular modular covariance of
the observables and implies, as mentioned in the introduction, the
\CPT theorem on the observable level. 
(Of course for these properties to hold true,
asymptotic completeness on the {\em vacuum} sector is a sufficient
condition, namely $\calH^{(\Out)}\cap \calH_\iota=\calH_\iota$.)  
\section{The \CPT Theorem} \label{secCPT}
We now prove that the pseudo-modular conjugation $\JF$
is a \CPT operator on the level of the field algebra, namely 
satisfies  Eq.s~\eqref{eqCPT}, \eqref{eqCPTgeo} and \eqref{eqC}. 
As mentioned in the introduction, Guido and Longo have
shown~\cite{GL95} that modular covariance of the observables, 
which we have established in Theorem~\ref{ModCov}, 
implies that the corresponding modular conjugation $\JO$ 
is a ``PT'' operator on the 
observable level, namely satisfies 
Eq.~\eqref{eqCPT} on the vacuum Hilbert space and Eq.~\eqref{eqCPTgeo}
with $\calA(\spc)$ instead of $\F(\spc)$. 
As a consequence, the anti-isomorphism
$\alphaO_j:\Auni(W_1)\to\Auni(W_1')$ implemented by $\JO$,
cf.\ Eq.~\eqref{eqJ0alphaj}, extends to an anti-automorphism of the 
entire universal algebra $\Auni$ (still implemented by $\JO$). 
In particular, we have 
\begin{Cor}[\cite{GL95}]\label{CovConjObs} %
The modular conjugation $\JO$ of the observable algebra associated
with the wedge $W_1$ 
implements an anti-automorphism $\alphaO_j$ of the universal algebra $\Auni$,
\begin{equation} \label{eqJOImplement}
\Ad \JO\circ \pi_0 = \pi_0\circ \alphaO_j, 
\end{equation}
which has the representation properties
$\alphaO_j\alphaO_\po \alphaO_j= \alphaO_{j\po j}$, $\alphaO_j^2 = \iota$, 
and acts geometrically correctly:
\begin{equation} \label{eqAlphaj0Geo}
\alphaO_j :\Auni(\ccc)\rightarrow \Auni(j\ccc), \quad \ccc\in\Ccc.
\end{equation}
\end{Cor}
Guido and Longo also show that $\alphaO_j$ intertwines any localized 
morphism $\rho\in\Sec$ with its conjugate $\barrho$ up to equivalence: 
\begin{equation} \label{eqjRhoj}
\alphaO_j \circ \rho \circ \alphaO_j \simeq \barrho.
\end{equation}
(In fact, we exhibit an intertwiner establishing this equivalence in
Lemma~\ref{JOJ} below.)  
In order to make the \CPT theorem more explicit on the level of the
field algebra,
we begin with establishing the representation and implementation 
properties of the relative modular conjugations $J_\rho$. 
\begin{Prop} \label{JgJ}
The relative modular conjugations $J_\rho$, $\rho\in\Sec$, 
represent the reflection
$\J$ in the direct sum representation $U_\rho\oplus U_\barrho$, 
and  implement $\alphaO_j$ in the direct sum representation 
$\pi_0\rho\oplus\pi_0\barrho$ of  $\Auni$. Namely, there holds 
\begin{align} \label{eqJRepRho} 
J_\rho U_\rho(\potild) J_\rho^{-1} & = U_\barrho(j\potild j),\quad
\potild\in\Potild ,\\
\Ad J_\rho \circ \pi_0\rho & =\pi_0\barrho\circ \alphaO_j \quad
\text{on } \Auni. \label{eqJImplement}
\end{align} 
\end{Prop}
Of course, Eq.~\eqref{eqJRepRho} implies that also the pseudo-modular
conjugation $\JF$ represents the reflection, $\JF U(\potild) \JF^{-1}
= U(j\potild j)$. 
\begin{Proof} 
Eq.~\eqref{eqJRepRho} corresponds to Prop.~2.8 of Guido and Longo's
article \cite{GL95}, whose proof uses Borchers' commutation relations, 
commutation of the modular 
unitary group with the modular conjugation, and an
assertion~\cite[Thm.\ 1.1]{GL95} about unitary representations of the
universal covering group of $SL(2,\RR)$. But the commutation relations 
also hold in the present setting, c.f.\ Eq.s~\eqref{eqDUD} through
\eqref{eqJJ}, and the universal covering group of $SL(2,\RR)$
is just the homogeneous part of our $\Potild$,  
hence the proof of Guido and Longo goes through in the present case. 
The implementing property~\eqref{eqJImplement} has been shown
in~\cite[Prop.~1]{Mu_BorchersCR} to hold on $\Auni(W_1)$. By 
Eq.~\eqref{eqJRepRho} it extends to all wedges $W=\po W_1$. 
Considering a space-like cone $\spc$, we note that $\spc$ coincides with the 
intersection of the wedge regions containing $\spc$. Then Haag duality
for wedges and for space-like cones implies that 
\begin{equation*} 
\Auni(\spc) = \bigcap_{W\supset \spc} \Auni(W). 
\end{equation*}
(This can be seen by  the arguments in the proof of Cor.~3.5 in~\cite{BGL}.)  
This implies that Eq.~\eqref{eqJImplement} also
holds for space-like cones, and further, again by Haag duality, for their
causal complements. Thus Eq.~\eqref{eqJImplement} holds for all
$\ccc\in\Ccc$, and the proof is complete. 
\end{Proof}

Before calculating  the adjoint action of $\JF$ on the fields, we discuss the 
item of geometrical correctness~\eqref{eqCPTgeo}. In order to formulate 
it, one needs
an action of $\J$ on the set $\Cccpaths$ of paths of space-like cones,
such that $\J\act\cccpath$ is a path which starts at $\spc_0$ and ends
at $\J \ccc$. 
Since $\J$ cannot be continuously transformed to the identity
transformation, such action must be of the form 
$$
\J\act (C_0,I_1,\ldots,I_n) = (C_0,I,\J \spc_0,\J I_1,\cdots,\J I_n),
$$
where $I\in\Ccc$ contains both $C_0$ and $\J C_0$. 
Now our requirement that $\spc_0$ be contained in $W_1$ implies that 
$\spc_0$ and $\J\spc_0$ are causally separated. Hence there are
two topologically distinct regions $I^\pm$ containing $\spc_0\cup
\J\spc_0$, and the action of $\J$ on $\Cccpaths$ depends on this choice. 
The action would be canonical, however,  if the reference cone 
$\spc_0$ (which enters in the definition of $\Cccpaths$) were
invariant 
under $j$.\footnote{Again, there 
are two topologically distinct possibilities for the choice of
$\spc_0$: It must contain either the positive or of
the negative $x^2$-direction. 
The difference between the two choices shows up, in our context, only
in the action of the twist operator for Anyons, see
Eq.~\eqref{eqTwistExpl}. So far, we leave the choice of $\spc_0$
unspecified.} 
Namely, then $\J$ would
act canonically as  
\begin{equation} \label{eqjAct}
j\act (\spc_0,I_1,\ldots, I_n) := (j\spc_0, j I_1,\ldots, j I_n).
\end{equation}
We therefore wish to go over to such a reference cone. To this end 
we first convince ourselves that all relevant structure of the field
algebra is preserved under such change. More precisely, we show that 
two field algebras constructed as in 
Section~\ref{secAss} from the same observables and 
sectors, but with different reference cones, are isomorphic in
all structural elements mentioned in Section~\ref{secAss}:  
\begin{Lem} \label{FARef}
Let $\hat\Sec$ be a collection of morphisms localized in a
reference cone $\hat \spc_0$ (instead of $\spc_0$), one per 
equivalence class, and let 
$\hat \F$, $\widehat\Cccpaths$, $\hat\F(\cccpath)$ for  
$\cccpath\in\widehat\Cccpaths$, $\hat\calH$ and $\hat U(\potild)$ 
be defined as in Section~\ref{secAss} with $\Sec$ replaced by $\hat\Sec$. 
Then there is a bijection $\cccpath\mapsto \hat\cccpath$ from 
$\Cccpaths$ onto $\hat \Cccpaths$ and an isometric isomorphism 
$W:\calH\rightarrow \hat\calH$  
which implements a unitary equivalence $\F\cong\hat \F$ preserving the 
respective notions of localization and pseudo-adjoints, and intertwining the 
representations $U$ and $\hat U$ as well as the vacua. In formulas: 
\begin{align} 
\Ad W: \F(\cccpath) &\to \hat\F(\hat\cccpath),
                       \label{eqHatFC}&\cccpath&\in\Cccpaths,  \\
W F^\adj W^* & = (WFW^*)^\adj,       & F & \in \F_\iota,\label{eqHatFAdj} \\
W\, U(\potild)\, W^* &= \hat U(\potild), &\potild &\in
\Potild,  \label{eqHatU}  \\
W\Om & = \hat \Om.
\end{align}
\end{Lem}
\begin{Proof}
We choose $\ccc_0\in\Ccc$ which contains both $\spc_0$ and 
$\hat\spc_0$, and pick for each $\rho\in\Sec$ with corresponding
$\hat\rho \in \hat\Sec$ a unitary intertwiner
$W_\rho\in \Auni(\ccc_0)$ such that $\hat \rho = \Ad W_\rho\circ \rho$. 
Note that $\hat\barrho$ coincides with the conjugate of $\hat\rho$,
since $\Sec$ and $\hat\Sec$ contain exactly one morphism per class. 
As intertwiner from $\iota$ to $\bar{\hat{\rho}}\hat\rho$ we choose 
$$
R_{\hat\rho} := W_{\barrho} \,\barrho(W_\rho)\, R_\rho\equiv
\hat\barrho(W_\rho)\, W_\barrho \,R_\rho.
$$
For $e=(\rho_s,\rho_c,\rho_r,i)$ let 
$\hat e:=(\hat\rho_s,\hat\rho_c,\hat\rho_r,i)$ and 
$$
T_{\hat e} := (W_{\rho_s}\times W_{\rho_c})\, T_e \, W_{\rho_r}^* 
\equiv W_{\rho_s} \rho_s(W_{\rho_c}) \, T_e \, W_{\rho_r}^*
=\hat\rho_s(W_{\rho_c}) W_{\rho_s} \, T_e \, W_{\rho_r}^*.
$$
If $i$ runs through $1,\ldots,$ dim $\Int(\rho_s\rho_c|\rho_r)$, then
the $T_{\hat e}$ are an orthonormal basis of
$\Int(\hat\rho_s\hat\rho_c|\hat\rho_r)$. 
Let finally $W:\calH\to\hat\calH$ be defined by 
$W(\rho,\psi) := (\hat \rho,\pi_0(W_\rho) \psi)$. 
Then there holds 
$W \, F(e,A)\, W^* = \hat F(\hat e, W_{\rho}\,A)$,  
where $\rho$ is the charge of $e$.
Defining a bijection $\Cccpaths\to\hat\Cccpaths$ by 
$$
(\spc_0,\ccc_1,\ldots,,\ccc_n)=\cccpath \mapsto 
\hat \cccpath = (\hat \spc_0,I_0,\spc_0,\ccc_1,\ldots,\ccc_n),  
$$ 
one also checks that $F(e,A)\in\F(\cccpath)$ if and only if
$\hat F(\hat e,A)\in\hat\F(\hat\cccpath)$. 
Eq.~\eqref{eqHatFAdj} is easily verified, and the intertwiner 
relation~\eqref{eqHatU} follows from~\cite[Lem.~2.2]{DHRIV}. 
\end{Proof}
The lemma implies of course that the pseudo-modular objects of 
$\hat\F$ have the same algebraic relations among themselves and with 
the representation $\hat U(\Potild)$ as those of $\F$. Namely, it
implies that the new pseudo-Tomita operator $\hat \SF$ defined via 
\begin{equation*} 
\hat S\, \hat F\hat\Om := \hat F^\adj\hat \Om, \quad \hat
F\in\hat\F(\hat{\tilde{W}}_1), 
\end{equation*}
satisfies $\hat S = W\, \SF\, W^*$, and we have the following 
\begin{Cor} 
All commutation relations between the relative modular objects and the
Poincar\'e transformations, namely Eq.s 
{\rm (\ref{eqJDJ}, \ref{eqJJ}, \ref{eqModCov}, 
\ref{eqJRepRho})}, as well as the implementation 
property~\eqref{eqJImplement}, 
also hold with $\rho,\Delta_\rho$, $J_\rho$, $U_\rho$ replaced by 
$\hat\rho,\hat \Delta_{\hat\rho}$, $\hat J_{\hat\rho}$,
$U_{\hat\rho}$, respectively.   
\end{Cor}
Here, $ \hat J_{\hat\rho}$ and $\hat\Delta_{\hat\rho}$ are defined
from $\hat \SF\equiv\hat \JF\,\hat \Delta^{1/2}$ as in
Eq.s~\eqref{eqModRho}, \eqref{eqModRho'}. 
\begin{Proof}
Only the implementation property~\eqref{eqJImplement} of the new
conjugation $\hat J_{\hat\rho}$ remains to be shown. But this property
follows from the fact that $W_\rho$ intertwines the morphisms
$\hat\rho$  and $\rho$ by  construction, c.f.\ the proof of Lemma~\ref{FARef}.
\end{Proof}
Due to this result, we may from now on assume that the reference cone 
$\spc_0$ in which all morphisms $\rho\in \Sec$ are localized is 
invariant under $\J$, $\J\spc_0 =\spc_0$, and that $\J$ acts on 
$\Cccpaths$ in a canonical way as in Eq.~\eqref{eqjAct}. 
We now wish to calculate the action of $\JF$ on the field algebra, and
to this end introduce ``cocycle'' type unitaries $V_\rho$
which  implement the equivalence~\eqref{eqjRhoj} of $\alphaO_j 
\rho \alphaO_j$ and $\barrho$: 
\begin{Lem} \label{JOJ}
The unitary operator $J_\rho \JO$ is in $\AO(\spc_0)$. Its pre-image 
under the (faithful) restriction of $\pi_0$ to
$\Auni(\spc_0)$ is an intertwiner from $\alphaO_j
\rho \alphaO_j$ to $\barrho$.  In other words, the unitary 
\begin{equation}  \label{eqVRho}
V_\rho  := \pi_0^{-1} (J_\rho \JO )
\end{equation}
is in $\Auni(\spc_0)$ and satisfies 
\begin{equation}  \label{eqJOJInt} 
\Ad V_\rho \circ \alphaO_j\, \rho\, \alphaO_j  = \barrho. 
\end{equation}
\end{Lem}
\begin{Proof} 
The implementation properties  \eqref{eqJOImplement} and 
\eqref{eqJImplement} imply that 
\begin{equation*} 
\Ad (J_\rho\JO ) \circ \pi_0 \,\alphaO_j\rho\alphaO_j   =\pi_0 \barrho. 
\end{equation*}
The morphisms on the left and right hand sides are localized in
$\spc_0$ and $\J \spc_0$, respectively. 
Hence by Haag duality, $J_\rho\JO $ is in $\AO(I)=\pi_0\Auni(I)$,
whenever $I$ contains $\spc_0 \cup \J\spc_0$. 
The rest of the proof is straightforward.
\end{Proof}
Our explicit formula for the adjoint action of $\JF$ below  uses the
unitaries $V_\rho$ and relies on the following
observation. Recall that for each superselection channel 
$e=(\rho_s,\rho_c,\rho_r,i)$ we have chosen an intertwiner 
$T_e$ in $\Int(\rho_s\rho_c|\rho_r)$. One easily verifies that 
$$
\bar T_e := 
(V_{\rho_s}\times V_{\rho_c}) \,\alphaO_j(T_e)\,V_{\rho_r}^* 
\equiv \barrho_s(V_{\rho_c})V_{\rho_s}\alphaO_j(T_e) V_{\rho_r}^*  
$$
is an intertwiner in $\Int(\barrho_s\barrho_c|\barrho_r)$, and 
can therefore be expanded in terms of the basis $\{T_{\bar e}\}$, where 
$\bar e$ is of type $(\barrho_s,\barrho_c,\barrho_r)$.
Therefore, 
\begin{equation} \label{eqTTbar}
\bar T_e = \sum_{\bar e} c_{\bar e,e} \, T_{\bar e},\quad 
\text{ with }\quad  c_{\bar e,e}\,\unity:= T_{\bar e}^*\,\bar T_e,
\end{equation}
where the sum goes over all superselection channels $\bar e$ of 
type $(\barrho_s,{\barrho_c},{\barrho_r})$.
We are now prepared for the \CPT theorem. 
\begin{Thm}[\CPT Theorem.] \label{ModCovConj}
The pseudo-\-modular conjugation $J$ is a \CPT operator in the sense
of Eq.s~\eqref{eqCPT}, \eqref{eqCPTgeo} and \eqref{eqC}.
Namely, it represents the reflection $j$  in a geometrically
correct way: 
\begin{align} \label{eqJRep}
J\,U(\potild)J^{-1} &= U(j\potild j), \quad \potild\in\Potild, \\
\label{eqJGeo} 
\Ad \JF:\, \F(\cccpath)&\to \F(j\act \cccpath), \quad \cccpath\in\Cccpaths,
\end{align}
and conjugates charges. More explicitly, $\Ad\JF$ is given by 
\begin{equation} \label{eqJFExpl}
\JF\,F(e,A)\,\JF^{-1} = \sum_{\bar{e}} \bar c_{\bar e,e} \, F(\bar
e,V_{{\rho_c}}\alphaO_j(A)), 
\end{equation}
where the sum goes over all superselection channels $\bar e$ of 
type $(\barrho_s,{\barrho_c},{\barrho_r})$ if
$e$ is of type $(\rho_s,\rho_c,\rho_r)$. 
\end{Thm}
It is interesting to note the resemblance of Eq.s~\eqref{eqVRho} and 
\eqref{eqJFExpl} with Eq.s ~\eqref{eqYrho} and \eqref{eqUImplement},
respectively.  
\begin{Proof} 
The representation property~\eqref{eqJRep} has been shown in 
Proposition~\ref{JgJ}. To prove Eq.~\eqref{eqJFExpl}, let $e$ be of type
$(\rho_s,\rho_c,\rho_r)$. We have 
\begin{align*}
\JF F(e,A)\JF^{-1} (\barrho_s,\psi) 
&= \big(\barrho_r,\pi_0\big(V_{\rho_r}\alphaO_j(T_e^*)
 V_{\rho_s}^*\barrho_s\alphaO_j(A)\big)\psi\big)\\ 
& = \sum_{\bar e} \bar c_{\bar e,e}\, 
\big(\barrho_r,\pi_0\big(T_{\bar e}^*\barrho_s(V_{\rho_c}
\alphaO_j(A)\big)\psi\big) \\
&= \sum_{\bar e} \bar c_{\bar e,e}\, 
F(\bar e,V_{\rho_c}\alphaO_j(A))\,(\barrho_s,\psi).  
\end{align*}
In the second equality we have used that 
$V_{\rho_r}\alphaO_j(T_e^*) V_{\rho_s}^*=
(\bar T_{e})^* \barrho_s(V_{\rho_c})$ by definition of $\bar
T_e$, and substituted $\bar T_e$ as in Eq.~\eqref{eqTTbar}. 
This proves Eq.~\eqref{eqJFExpl}, and shows that $\Ad \JF$ is an
anti-automorphism of the field algebra. In order to show that it acts 
geometrically correctly, let 
$\cccpath=(I_0=\spc_0,I_1,\ldots,I_n=\ccc)\in\Cccpaths$, let 
$U=U_n\cdots U_1$ be a charge transporter for $\rho$ along
$\cccpath$, and let $F(e,A)$ be in $\F(\cccpath)$ with $c(e)=\rho$. 
We define   
$$
\bar U_k:=\alphaO_j(U_k),\;k=1,\ldots,n. 
$$
Then $\bar U:= \bar U_n\cdots \bar U_1V_{\rho}^*$ is a charge transporter for
$\barrho$ along $j\act \cccpath$, since 
$$ 
\Ad (\bar U_k\cdots \bar U_1V_{\rho}^*)\circ \barrho \equiv 
\alphaO_j\circ \Ad (U_k\cdots U_1)\circ \rho\circ \alphaO_j 
$$ 
is localized in $j I_k$, $k=1,\ldots,n$. Further, 
$\bar U V_\rho \alphaO_j(A) \equiv \alphaO_j(UA)$ 
is in $\Auni(j\ccc)$ by Eq.~\eqref{eqAlphaj0Geo}, since by hypothesis  
$UA\in\Auni(\ccc)$. Hence 
$F(\bar e,V_{\rho_c}\alphaO_j(A))$ is in $\F(\J\act \cccpath)$, and
the proof is complete. 
\end{Proof}

\section{Anyons} \label{secAnyons}
We now consider the case of Anyons, i.e., when all sectors correspond to
{automorphisms} of the observable algebra. 
In this case,  one can construct a 
field algebra $\FA$~\cite{Re90,MDiss,S94} in the sense of the WWW
scenario, in particular the  vacuum vector is
cyclic and separating for the local field algebras. 
Thus, the Tomita operator associated with $\FA(\We)$
and the vacuum is well-defined and one may ask whether there
holds modular covariance. We show that this is indeed the case. 
To this end, we exhibit $\FA$ as a sub-algebra of the reduced 
field bundle $\F$, and show that the pseudo-modular operator of $\F(\We)$
coincides with the (genuine) Tomita operator of $\FA(\We)$. Then
modular covariance is implied by  the results from the previous
sections, and the \CPT theorem follows in analogy with the permutation
group statistics case. 

We assume for simplicity that the set $\SecE$ of elementary charges 
consists of only one
automorphism, localized in $\spc_0=\J \spc_0$.  (All results
are easily transferred to the case of finitely many elementary Abelian
charges, i.e., $\SecE$ finite.)  
$\Sec$ has then the structure of an Abelian group with one generator,
i.e., $\II_N$ if there is a natural number $N$ such that 
the $N$-fold product of the generating automorphism is equivalent to the
identity, and  $\II$ otherwise. 
In the former case there may or may not be a representant whose
$N$-fold product coincides with the identity, as Rehren has 
pointed out~\cite{Re90}. 
We wish to exclude the case with obstruction for
simplicity, and therefore make the following
\begin{Ass} \label{A3}
The set of relevant sectors $\Sec$ is generated by one automorphism,
$\gamma$. 
If there is a natural number $N$ such that 
$\gamma^N:=\gamma\circ\cdots\circ\gamma$ is equivalent to
the identity $\iota$, then $\gamma$ can be chosen such that $\gamma^N=\iota$. 
\end{Ass}
\subsection{The Field Algebra for Anyons.} 
In the assumed absence of an obstruction, there is a field 
algebra $\FA$ in the sense of the WWW 
scenario as mentioned in the introduction: 
The dual group of $\Sec$, in our case $U(1)$ or $\II_N$, acts
as a global gauge group on the local algebras, singling out the local
observables as invariants under this action. The vacuum vector is
cyclic and separating for the local field algebras. Further, the local 
commutation relations (governed by an Abelian representation of the 
braid group) 
can be formulated in terms of twisted locality~\eqref{eqTwistedLoc}, as
in the familiar case of Fermions~\cite{DHRI}.\footnote{Some results on 
 the anyonic field algebra in $d=2+1$ are spread out over the 
literature, or have not been made very explicit in the accessible 
literature (see however~\cite{MDiss}), namely: Local 
 commutation relations under consideration of the relative winding
 numbers, twisted duality, and the Reeh-Schlieder property. We therefore
 collect them, with proofs, in the appendix for the convenience of the
 reader.}  
The field algebra $\FA$ for Anyons~\cite{Re,MDiss} 
is constructed as follows. 
Due to our Assumption~\ref{A3}, the map 
$$ 
\gamma^q\mapsto q, \quad q\in\II_N \text{ or } \II 
$$
establishes an isomorphism of the groups $\Sec$ and $\II_N$ or $\II$,
respectively. 
We shall identify $\Sec$ with $\II_N$ or $\II$ by this isomorphism, and write
$q$ instead of ${\gamma^q}$, $\calH_q$ and $E_q$ instead of
$\calH_{\gamma^{q}}$ and $E_{\gamma^q}$, and so on. 
We consider the  same Hilbert space $\calH$  and 
representation $U$ of the universal covering group of the Poincar\'e group as
before, see Eq.s~\eqref{eqHS} and \eqref{eqU}. 
The dual $\hat \Sec$ of the group $\Sec$ is called the gauge group and
is represented on $\calH$ via 
\begin{equation} \label{eqGaugeGroup} 
V(t):=\sum_{q\in\Sec} \exp(2\pi i qt)\,E_q.  
\end{equation}
The anyonic field algebra $\FA$ is now the $C^*$-algebra generated by 
operators $\fa(c,A)$, $c\in\Sec$, $A\in\Auni$, acting as 
\begin{equation*} 
\fa(c,A): \;  (q,\psi) \mapsto (q+c,\pi_q(A)\psi). 
\end{equation*}
Here we have written $\pi_q:= \pi_0\circ\gamma^q$ to save on notation. 
Clearly there holds 
$\fa(c_1,A_1)\fa(c_2,A_2)=\fa(c_1+c_2,\pi_{c_2}(A_1)A_2)$,
 and therefore $\FA$ coincides with the closed linear span of the
 operators $\fa(c,A)$, $A\in \Auni$, $c\in \Sec$.  
Furthermore, the adjoint is given by 
\begin{equation}  \label{eqAdjAny}
\fa(c,A)^* = \fa\big(-c,\gamma^{-c}(A^*)\big). 
\end{equation}
As in the case of the reduced field bundle, we call a field operator 
$\fa(c,A)$ localized in
$\cccpath\in\Cccpaths$ if there is a charge transporter $U$ for
$\gamma^c$ along $\cccpath$ such that $UA\in\Auni(\ccc)$. 
The von Neumann algebra generated by these
operators is denoted by $\FA(\cccpath)$. 
The adjoint action of the gauge group~\eqref{eqGaugeGroup} leaves each
local field algebra $\FA(\cccpath)$ invariant, the fixed point algebra 
being the direct sum of all $\pi_q(\Auni(\ccc))$, hence isomorphic to
$\Auni(\ccc)$. 
The space-like commutation relations have the following explicit
form. 
Let $F_1=\fa(c_1,A_1)\in\FA(\cccpath_1)$ and 
$F_2=\fa(c_2,A_2)\in\FA(\cccpath_2)$, where 
$\cccpath_1,\cccpath_2$ are causally separated and have relative
winding number $N(\cccpath_2,\cccpath_1)=n$. Then there
holds\footnote{See appendix.\label{Anyons}}  
\begin{equation*} 
 F_2\, F_1  = \omega^{c_1\cdot c_2(2n+1)} \,F_1\, F_2,
\end{equation*} 
where $\omega\equiv\omega_\gamma$ denotes the statistics phase of the 
generating 
automorphism $\gamma$, see~Eq.~\eqref{eqCommut}.  
This may be reformulated in the form of twisted locality, which by
Haag duality~\eqref{eqHD} sharpens to twisted duality, as follows. 
{}For $\cccpath_1$, $\cccpath_2$ causally separated with relative
winding number $N(\cccpath_2,\cccpath_1)=n$, let 
$Z(\cccpath_2,\cccpath_1)$ be the unitary operator in $\calH$ defined  by 
\begin{equation} \label{eqTwistOp} 
Z(\cccpath_2,\cccpath_1) \; E_q := \big(\omega^{\frac{1}{2}}\big)^{ 
q^2(2n+1)} \; E_q, 
\end{equation}
where the root of $\omega$ may be chosen at will. 
(This ``twist operator'' has been first proposed in~\cite{S94}.) 
\begin{Lem}[Twisted Haag Duality.] 
Let $\cccpath, \cccpath'$ be (classes of) paths in $\Cccpaths$ ending
at $\ccc$ and its causal complement $\ccc'$, respectively. Then there holds
\begin{equation} \label{eqTwistedDual} 
Z(\cccpath,\cccpath')\, \FA(\cccpath')\,
Z(\cccpath,\cccpath')^*  =  \FA(\cccpath)'. 
\end{equation}
\end{Lem}
(We give a proof of this lemma in the appendix.)
\subsection{Modular Covariance and CPT Theorem.} 
We now show that the anyonic field algebra satisfies covariance of the 
modular groups and conjugations. 
Since the vacuum is cyclic and separating for the local algebras, see
Lemma~\ref{ReehSchlieder}, the Tomita 
operator associated with $\FA(\We)$ is well-defined. Let us denote
this operator and its polar decomposition by 
$$
\SA=\JA\,\DA^{1/2}.  
$$ 
\begin{Thm}[Modular Covariance.]     \label{ModCovAny} 
Let the Assumptions~{\rm \ref{A1}}, {\rm \ref{A2}} and {\rm \ref{A3}} 
hold. Then the modular unitary group of the anyonic field algebra satisfies 
modular covariance, namely coincides with the representers of the 1-boosts: 
\begin{equation*} 
\DA^{it}= U(\boox{-2\pi t}).
\end{equation*}
\end{Thm}
\begin{Proof}
The theorem is a simple consequence of our Theorem~\ref{ModCov} and
the following lemma. 
\end{Proof}
\begin{Lem} \label{SFFA}
The anyonic field algebra $\FA$ is a sub-algebra of the reduced field
bundle $\F$, and the pseudo-Tomita operator $\SF$ associated with 
$\F(\We)$ and $\Om$
coincides with the Tomita operator $\SA$ associated with $\FA(\We)$ and $\Om$. 
\end{Lem}
(The fact that then $\SF$ must be an involution, $\SF^2\subset\unity$, 
is no contradiction to Eq.~\eqref{eqS2}, since if $\gamma$ is 
self-conjugate our Assumption~\ref{A3} implies that 
$\gamma$ is real and not pseudo-real~\cite[Remark 2 after
Lem.~4.5]{Re90}, hence $\chi_\gamma=1$.) 
\begin{Proof} 
Let us first set up the reduced field bundle in the special case at
hand. 
We identify $\Sec$ with $\II$ or $\II_N$ as before, and
denote elements by $q,s,c,r$. 
A superselection channel  $e=(s,c,r)$ has non-zero intertwiner 
$T_e\in\Int(s+c|r)$ only if $r=s+c$. In this case we choose
$T_e=\unity$ and write $F(s,c;A)$ instead of $F(e,A)$.  
The field algebra $\F$ (alias reduced field bundle) is thus generated by
the operators 
\begin{equation} \label{eqRedFieldAny}
F(s,c;A): \;  (q,\psi) \mapsto \delta_{s,c}\; (q+c,\pi_q(A)\psi), 
\end{equation}
$s,c\in\Sec$, $A\in\Auni$.
Clearly, $\fa(c,A)=\sum_{s\in\Sec}\,F(s,c;A)$, and hence $\FA$ is a
sub-algebra of $\F$. Further, the map\footnote{In fact, $\CE$
  is a conditional expectation from $\F$ onto $\FA$.}  
$\CE:\F\mapsto \FA$ defined by 
\begin{equation} \label{eqCE} 
\CE \big( F(s,c;A)):= \fa(c,A)
\end{equation} 
preserves localization, and for $F\in \F_\iota$ satisfies 
 $\CE(F)\,\Om  = F\,\Om$ and $\CE\big(F^\adj\big)  = \CE(F)^*$. 
(The last equation follows from  
$$
F(0,c;A)^\adj = F\big(0,-c;\gamma^{-c}(A^*)\big) 
$$ 
and Eq.~\eqref{eqAdjAny}.) 
These relations imply that for $F\in\F_\iota$ there holds 
$$
S\,\CE(F)\Om = S\,F\Om = F^\adj\Om=\CE(F^\adj)\Om=\CE(F)^*\Om.
$$ 
But $\CE$ clearly maps $\F(\We)$ {\em onto} $\FA(\We)$, 
and the proof is complete. 
\end{Proof}
The lemma also implies, of course, that the  modular conjugation $\JA$ 
associated with $\FA(\We)$ represents the reflection $\J$ in the sense 
of Eq.~\eqref{eqCPT} or \eqref{eqJRep}. 
In order to achieve a  geometrically correct action on the family of algebras
$\FA(\cccpath)$ in the sense of Eq.~\eqref{eqJGeo}, however, one has
to multiply it with the twist operator $Z$, 
\begin{equation} \label{eqTwist}
Z:= Z(\We,\J\act\We). 
\end{equation}
\begin{Thm}[\CPT Theorem for Anyons.] \label{CPTAny}
The anti-unitary operator 
$
\CPTop := Z^*\,\JA 
$ 
is a \CPT operator: It 
satisfies 
\begin{equation} \label{eqCPTRep}
\CPTop^2 = \unity,\qquad \CPTop U(\potild)\CPTop^{*} = U(j\potild j), 
\end{equation}
$\potild \in\Potild$, and acts geometrically 
correctly: For all $\cccpath\in\Cccpaths$, there holds 
\begin{equation} \label{eqCPTGeo}
\Ad \CPTop : \; \FA(\cccpath) \to  \FA(j\act\cccpath). 
\end{equation}
\end{Thm}
\begin{Proof}
The commutation relations Eq.~\eqref{eqCPTRep} of $\CPTop$ are
inherited from  those of $\JA\equiv \JF$, since $Z$ commutes with
$U(\Potild)$. $\JA^2=\unity$, anti-linearity of $\JA$ and $\JA E_q=E_{-q} \JA$ 
imply that $\CPTop^2=\unity$. 
Tomita-Takesaki's theorem and twisted duality~\eqref{eqTwistedDual} imply that 
$$
\JA\,\FA(\We)\,\JA^* = \FA(\We)'= Z\,\FA(\J\act \We)\,Z^*,
$$
which yields Eq.~\eqref{eqCPTGeo} for the case of $\We$. Covariance
then implies the geometric action for every $\cccpath\in\Cccpaths$
which ends at a wedge region $W$, namely, every $\cccpath$ of the form
$\potild\act \We$. If $\cccpath$ ends at a space-like cone, 
note that twisted Haag duality~\eqref{eqTwistedDual} implies that 
$\FA$ is self-dual, namely 
$$
\FA(\cccpath)=\bigcap_{\tilde{W}\supset \cccpath}\,\FA(\tilde{W}), 
$$
where the intersection goes over all $\tilde{W}\in\Cccpaths$ which
contain $\cccpath$. (By $\cccpath\subset\tilde{W}$  we mean that 
$\ccc\subset W$ and in addition $\cccpath^H\subset \tilde{W}^H$ as 
subsets of $\tilde H$.)   
Hence Eq.~\eqref{eqCPTGeo} also  holds for such $\cccpath$.  
If $\cccpath$ ends at the causal complement of a space-like cone,  
the equation also holds, since $\FA(\cccpath)$ is generated by all 
$\FA(\spcpath)$ with $\spcpath\subset\cccpath$. (This is so because the
analogous statement holds for $\Auni(I)$.)  
This completes the proof of the theorem. 
\end{Proof}
Up to here, the reference cone $\spc_0$ has not been specified, and
$\We$ may be any path from $\spc_0$ to $W_1$. The difference in the
choices only shows up in the value of the twist operator $Z$. 
Recall that there are two topologically distinct choices for $\spc_0$ 
satisfying $\J \spc_0=\spc_0$. With any one of these choices, the
natural choice of $\We$ is the ``shortest'' path from $\spc_0$
to $W_1$, namely 
$$ 
\We := (\spc_0,I,W_1) 
$$
for some $I\in\Ccc$. 
Specifying now $\spc_0$ so as to contain the positive or negative $x^2$-axis,
respectively, the 
relative winding number of $\We$ and $\J\act \We$ is 
$N(\We,\J\act \We) = -1$ or $0$, respectively, hence 
\begin{equation} \label{eqTwistExpl} 
Z\, E_q = \omega^{\mp\half q^2} \, E_q, 
\end{equation}
respectively. 
\appendix
\section{The Anyon Field Algebra} 
We collect some results on the anyonic field algebra which
are spread out over the literature, or have not been made very
explicit in the literature. 
\begin{Lem}[Anyonic Commutation Relations.] \label{AnyCR} 
Let $\cccpath_1,\cccpath_2$ be causally separated, with relative
winding number $N(\cccpath_2,\cccpath_1)=n$. 
\\
i) For $F_1=\fa(c_1,A_1)\in\FA(\cccpath_1)$ and 
$F_2=\fa(c_2,A_2)\in\FA(\cccpath_2)$
there hold the commutation relations 
\begin{equation}\label{eqCommutAny}
 F_2\, F_1  = \omega^{c_1\cdot c_2(2n+1)} \,F_1\, F_2,
\end{equation} 
where $\omega$ denotes the statistics phase of the generating 
automorphism $\gamma$, see~Eq.~\eqref{eqCommut}.  
\\
ii) Equivalent with these relations is {\em twisted locality}, namely 
\begin{equation} \label{eqTwistedLoc} 
Z(\cccpath_2,\cccpath_1)\, \FA(\cccpath_1)\,
Z(\cccpath_2,\cccpath_1)^*  \subset \FA(\cccpath_2)'
\end{equation}
if $\ccc_1$ and $\ccc_2$ are causally separated. 
\end{Lem}
Here, $Z(\cccpath_2,\cccpath_1)$ is the ``twist'' operator defined in
Eq.~\eqref{eqTwistOp}.  
\begin{Proof}  
Ad $i)$ The commutation relations~\eqref{eqFieldCR} satisfied by the 
{\em reduced} field operators read as follows in the present context 
of Anyons.  
Two fields $F(s_1,c_1;A_1)\in\F(\cccpath_1)$ and 
$F(s_2,c_2;A_2)\in\F(\cccpath_2)$, where $s_2=s_1+c_1$, which are
causally separated satisfy the commutation relations
\begin{equation} \label{eqFieldCRAny}
F(s_2,c_2;A_2)\,F(s_1,c_1;A_1)  = 
R(s_1,c_1,c_2;n) 
\,F(\hat s_1,c_1;A_1)\,F(s_1,c_2;A_2),  
\end{equation}
where $\hat s_1=s_1+c_2$, and where $n$ is the relative winding number
$N(\cccpath_2,\cccpath_1)$. The number 
$R(s_1,c_1,c_2;n)$ is given by, see Eq.~\eqref{eqRMat},  
\begin{equation} \label{eqRn}
R(s_1,c_1,c_2;n)= 
\big(\frac{\omega_\alpha\omega_\gamma}{\omega_\beta\omega_\delta}\big)^{n}\;
\pi_0\gamma^{s_1}
\big(\eps(\gamma^{c_2},\gamma^{c_1})\big)
\end{equation}
with $\alpha=s_1,\beta=s_2\equiv s_1+c_1$, $\gamma=s_2+c_2\equiv
s_1+c_1+c_2$, $\delta=\hat s_1\equiv s_1+c_2$. 
Now the statistics operator $\eps(\gamma^{c_2},\gamma^{c_1})$
coincides, in the present situation, with 
$\eps(\gamma,\gamma)^{c_1\cdot c_2}$~\cite[Eq.~(2.3)]{FRSII}. Further,
$\eps(\gamma,\gamma)$ coincides with a multiple $\omega\unity$ of
unity~\cite{DHRIII}, where $\omega\equiv\omega_\gamma$ is the
statistics phase of $\gamma$. Putting all this into Eq.~\eqref{eqRn} yields
$R(s_1,c_1,c_2;n)=\omega^{c_1c_2(2n+1)}$. 
Using this equality, 
the commutation 
relations~\eqref{eqFieldCRAny} transfer to the fields 
$\fa(c_i,A_i)\equiv \sum_sF(s,c_i;A_i)$, proving the claim.
Ad $ii)$ For $F_1, F_2$ as in $(i)$ and
$Z:=Z(\cccpath_2,\cccpath_1)$ one calculates 
$$ 
[F_2,ZF_1Z^*]\, E_q=\omega^{\half c_1^2 +q c_1}\big(
F_2F_1-\omega^{c_1c_2}F_1F_2\big)\,E_q.  
$$
Hence the anyonic commutation relations~\eqref{eqCommutAny} are
equivalent with $[F_2,ZF_1Z^*]=0$. This implies
that $ZF_1Z^*$ also commutes with the von Neumann algebra generated
by operators of the form $F_2$, and completes the proof.  
\end{Proof}
\begin{Lem}[Twisted Haag Duality.]  \label{TwistedDual} 
Let $\cccpath, \cccpath'$ be (classes of) paths in $\Cccpaths$ ending
at $\ccc$ and its causal complement $\ccc'$, respectively. Then there holds
\begin{equation} \label{eqTwistedDual'} 
Z(\cccpath,\cccpath')\, \FA(\cccpath')\,
Z(\cccpath,\cccpath')^*  =  \FA(\cccpath)'. 
\end{equation}
\end{Lem}
\begin{Proof}
This follows from twisted locality and Haag duality of the observables 
as in the 
permutation group case~\cite[Thm.~5.4]{DR90}, the 
argument being as follows in the present setting. 
A standard argument~\cite[Remark~1 after Prop.~2.2]{DR72} using the 
Reeh-Schlieder property and the action~\eqref{eqGaugeGroup} of the
gauge group implies that every operator $B$ in
$\FA(\cccpath)'$ decomposes, just like a field operator, as the sum
of operators $B_q\in\FA(\cccpath)'$ carrying fixed charge, i.e.\ 
$B_cE_q=E_{q+c}B_c$. (Namely, $B_c=\int_{\hat\Sec}dt \exp(-2\pi i ct)
V(t)BV(t)^*$.)  
The same holds for $Z^*\FA(\cccpath)'Z$, where 
$Z:=Z(\cccpath,\cccpath')$. Let now $F\in Z^*\FA(\cccpath)'Z$ carry  
charge $c$, and pick a unitary $\Psi\in\FA(\cccpath')$ of the same charge. 
Then $B:=\Psi^*F$ also is in 
$Z^*\FA(\cccpath)'Z$  by twisted locality, and carries charge
zero. Therefore it is in $\FA(\cccpath)'$, and acts according to 
$B \,(q,\psi)= (q,B_q\psi)$. 
One concludes that for every $q\in\Sec$, charge transporter $U_q$ 
for $\gamma^q$ along the path $\cccpath$ and
observable $A\in\Auni(\ccc)$ there holds 
$$ 
B_q \pi_0(U_q^*A) = \pi_0(U_q^*A) \, B_0.
$$ 
Putting $q=0$, this implies by Haag duality that $B_0=\pi_0(\hat B)$
for some $\hat B\in \Auni(I')$. 
The same equation (with $A=\unity$) then implies that $B_q =
\pi_0(U_q^*\hat B U_q)$, which coincides with $\pi_q(\hat
B)$. Thus, $B$ coincides with $\fa(0,\hat B)$ and is therefore in
$\FA(\cccpath')$, and the same holds for $F$. 
This completes the proof.
\end{Proof}
\begin{Lem}[Reeh-Schlieder Property.] \label{ReehSchlieder} 
The vacuum is cyclic and separating for every  $\FA(\cccpath)$,
$\cccpath\in \Cccpaths$. 
\end{Lem}
\begin{Proof}
Cyclicity of the vacuum $\Omega=(0,\Omega_0)$ for $\FA(\cccpath)$ follows from
the cyclicity of $\Omega_0$ for $\Auni(\ccc)$ and the 
definition $\fa(c,A)(0,\Omega_0)=(c,\pi_0(A)\Omega_0)$. 
Now by twisted locality~\eqref{eqTwistedLoc}, $\FA(\cccpath)'\Omega$
contains $Z\FA(\cccpath')\Omega$, with $Z$ unitary, which is
dense. Hence the vacuum is cyclic for $\FA(\cccpath)'$ and therefore
separating for $\FA(\cccpath)$. 
\end{Proof}
\paragraph{Acknowledgements.}
It is a pleasure for me to thank Klaus Fredenhagen, Daniele Guido, 
Roberto Longo and  Bernd Kuckert, to whose memory this article is
dedicated, for many stimulating discussions on the subject. 
\providecommand{\bysame}{\leavevmode\hbox to3em{\hrulefill}\thinspace}
\providecommand{\MR}{\relax\ifhmode\unskip\space\fi MR }
\providecommand{\MRhref}[2]{%
  \href{http://www.ams.org/mathscinet-getitem?mr=#1}{#2}
}
\providecommand{\href}[2]{#2}


\begin{thebibliography}{10}

\bibitem{Araki}
H.~Araki, \emph{Mathematical theory of quantum fields}, Int. Series of
  Monographs in Physics, no. 101, Oxford University Press, 1999.

\bibitem{BiWi}
J.J. Bisognano and E.H. Wichmann, \emph{On the duality condition for a
  {Hermitean} scalar field}, J. Math. Phys. \textbf{16} (1975), 985.

\bibitem{BiWi2}
\bysame, \emph{On the duality condition for quantum fields}, J. Math. Phys.
  \textbf{17} (1976), 303.

\bibitem{Borchers92}
H.J. Borchers, \emph{The {CPT}-theorem in two-dimensional theories of local
  observables}, Commun.\ Math.\ Phys. \textbf{143} (1992), 315--332.

\bibitem{Borchers98}
\bysame, \emph{On {P}oincar\'e transformations and the modular group of the
  algebra associated with a wedge}, Lett. Math. Phys. \textbf{46} (1998),
  295--301.

\bibitem{BBS}
H.J. Borchers, D.~Buchholz, and B.~Schroer, \emph{Polarization-free generators
  and the {S}-matrix}, Commun. Math. Phys. \textbf{219} (2001), 125--140.

\bibitem{BY00}
H.J. Borchers and J.~Yngvason, \emph{On the {PCT}-theorem in the theory of
  local observables}, Mathematical Physics in Mathematics and Physics (Siena)
  (R.~Longo, ed.), Fields Institute Communications, vol.~30, 2001, pp.~39--64.

\bibitem{BraRob}
O.~Bratteli and D.W. Robinson, \emph{Operator algebras and quantum statistical
  mechanics 1}, second ed., TMP, Springer, New York, 1987.

\bibitem{BGL93}
R.~Brunetti, D.~Guido, and R.~Longo, \emph{Modular structure and duality in
  conformal field theory}, Commun. Math. Phys. \textbf{156} (1993), 201--219.

\bibitem{BGL}
\bysame, \emph{Modular localization and {W}igner particles}, Rev.\ Math.\ Phys.
  \textbf{14} (2002), 759--786.

\bibitem{BuEp}
D.~Buchholz and H.~Epstein, \emph{Spin and statistics of quantum topological
  charges}, Fysica \textbf{17} (1985), 329--343.

\bibitem{BuF}
D.~Buchholz and K.~Fredenhagen, \emph{Locality and the structure of particle
  states}, Commun. Math. Phys \textbf{84} (1982), 1--54.

\bibitem{DHRI}
S.~Doplicher, R.~Haag, and J.E. Roberts, \emph{Fields, observables and gauge
  transformations {I}}, Commun. Math. Phys. \textbf{13} (1969), 1--23.

\bibitem{DHRIII}
\bysame, \emph{Local observables and particle statistics~{I}}, Commun. Math.
  Phys. \textbf{23} (1971), 199.

\bibitem{DHRIV}
\bysame, \emph{Local observables and particle statistics~{II}}, Commun. Math.
  Phys. \textbf{35} (1974), 49--85.

\bibitem{DR72}
S.~Doplicher and J.E. Roberts, \emph{Fields, statistics and non-{Abelian} gauge
  groups}, Commun. Math. Phys. \textbf{28} (1972), 331--348.

\bibitem{DR90}
\bysame, \emph{Why there is a field algebra with a compact gauge group
  describing the superselection structure in particle physics}, Commun. Math.
  Phys. \textbf{131} (1990), 51--107.

\bibitem{F81}
K.~Fredenhagen, \emph{On the existence of antiparticles}, Commun.~Math.~Phys.
  \textbf{79} (1981), 141--151.

\bibitem{F89}
\bysame, \emph{Structure of superselection sectors in low dimensional quantum
  field theory}, Proceedings (Lake Tahoe City) (L.L. Chau and W.~Nahm, eds.),
  1989.

\bibitem{F90}
\bysame, \emph{Generalizations of the theory of superselection sectors}, The
  algebraic theory of superselection sectors. Introduction and recent results.
  (D.~Kastler, ed.), World Scientific, 1990.

\bibitem{FGR}
K.~Fredenhagen, M.~Gaberdiel, and S.M. R\"{u}ger, \emph{Scattering states of
  plektons (particles with braid group statistics) in 2+1 dimensional field
  theory}, Commun. Math. Phys. \textbf{175} (1996), 319--355.

\bibitem{FRSI}
K.~Fredenhagen, K.-H. Rehren, and B.~Schroer, \emph{Superselection sectors with
  braid group statistics and exchange algebras {I}: General theory},
  Commun.~Math.~Phys. \textbf{125} (1989), 201--226.

\bibitem{FRSII}
\bysame, \emph{Superselection sectors with braid group statistics and exchange
  algebras {II}: Geometric aspects and conformal covariance}, Rev.~Math.~Phys.
  \textbf{SI1} (1992), 113--157.

\bibitem{FM1}
J.~Fr\"{o}hlich and P.A. Marchetti, \emph{Quantum field theories of vortices
  and anyons}, Commun. Math. Phys. \textbf{121} (1989), 177--223.

\bibitem{FM2}
\bysame, \emph{Spin-statistics theorem and scattering in planar quantum field
  theories with braid statistics}, Nucl. Phys. B \textbf{356} (1991), 533--573.

\bibitem{FroKer}
J.~Fr\"{o}lich and T.~Kerler, \emph{Quantum groups, quantum categories, and
  quantum field theory}, Lecture Notes in Mathematics, vol. 1542, Springer,
  Berlin, 1993.

\bibitem{Fuchs95}
J.~Fuchs, A.~Ganchev, and P.~Vecserny\'es, \emph{Rational {H}opf algebras:
  Polynomial equations, gauge fixing, and low dimensional examples},
  Int.~J.~Mod.~Phys, \textbf{A 10} (1995), 3431--3476.

\bibitem{GL92}
D.~Guido and R.~Longo, \emph{Relativistic invariance and charge conjugation in
  quantum field theory}, Commun. Math. Phys. \textbf{148} (1992), 521--551.

\bibitem{GL95}
\bysame, \emph{An algebraic spin and statistics theorem}, Commun. Math. Phys.
  \textbf{172} (1995), 517.

\bibitem{GL00}
\bysame, \emph{Natural energy bounds in quantum thermodynamics}, Commun. Math.
  Phys. \textbf{218} (2001), 513--536.

\bibitem{H96}
R.~Haag, \emph{Local quantum physics}, second ed., Texts and Monographs in
  Physics, Springer, Berlin, Heidelberg, 1996.

\bibitem{Hepp}
K.~Hepp, \emph{On the connection between {W}ightman and {LSZ} quantum field
  theory}, Axiomatic Field Theory (M.~Chretien and S.~Deser, eds.), Brandeis
  University Summer Institute in Theoretical Physics 1965, vol.~1, Gordon and
  Breach, 1966, pp.~135--246.

\bibitem{Jost}
R.~Jost, \emph{The general theory of quantized fields}, American Mathematical
  Society, Providence, Rhode Island, 1965.

\bibitem{Kuck}
B.~Kuckert, \emph{A new approach to spin \& statistics}, Lett. Math. Phys.
  \textbf{35} (1995), 319--331.

\bibitem{Kuck00}
\bysame, \emph{Two uniqueness results on the {U}nruh effect and on
  {PCT}-symmetry}, Commun. Math. Phys. \textbf{221} (2001), 77--100.

\bibitem{Longo97}
R.~Longo, \emph{An analogue of the {K}ac-{W}akimoto formula and black hole
  conditional entropy}, Commun.~Math.~Phys. \textbf{186} (1997), 451--479.

\bibitem{MackSchom}
G.~Mack and V.~Schomerus, \emph{Conformal field algebras with quantum symmetry
  from the theory of superselection sectors}, Commun.~Math.~Phys. \textbf{134}
  (1990), 139--196.

\bibitem{MDiss}
J.~Mund, \emph{{Quantum Field Theory of Particles with Braid Group Statistics
  in 2+1 Dimensions}}, Ph.D. thesis, Freie Universit\"at Berlin, 1998.

\bibitem{M01a}
\bysame, \emph{The {B}isognano-{W}ichmann theorem for massive theories}, Ann.
  H. Poinc. \textbf{2} (2001), 907--926.

\bibitem{M02a}
\bysame, \emph{Modular localization of massive particles with ``any'' spin in
  d=2+1}, J.\ Math.\ Phys. \textbf{44} (2003), 2037--2057.

\bibitem{Mu_BorchersCR}
\bysame, \emph{Borchers' commutation relations for sectors with braid group
  statistics in low dimensions}, to be published by Ann.\ H.\ Poinc., 2009.

\bibitem{Mu_SpiSta}
\bysame, \emph{The spin statistics theorem for anyons and plektons in d=2+1},
  Commun.\ Math.\ Phys. \textbf{286} (2009), 1159--1180.

\bibitem{MSY}
J.~Mund, B.~Schroer, and J.~Yngvason, \emph{String--localized quantum fields
  from {W}igner representations}, Phys.\ Lett.\ B \textbf{596} (2004),
  156--162.

\bibitem{ONeill}
B.~O'Neill, \emph{Semi--riemannian geometry}, Academic Press, New York, 1983.

\bibitem{PauliPCT}
W.~Pauli, \emph{Exclusion principle, {L}orentz group and reflection of
  space-time and charge}, Niels Bohr and the Development of Physics (W.~Pauli,
  ed.), Pergamon Press, 1955, p.~30.

\bibitem{Re90b}
K.-H. Rehren, \emph{Braid group statistics and their superselection rules}, The
  Algebraic Theory of Superselection Sectors (D.~Kastler, ed.), World
  Scientific, Singapore, 1990.

\bibitem{Re90}
\bysame, \emph{Spacetime fields and exchange fields}, Commun. Math. Phys.
  \textbf{132} (1990), 461--483.

\bibitem{Re}
\bysame, \emph{Field operators for anyons and plektons}, Commun. Math. Phys
  \textbf{145} (1992), 123.

\bibitem{Rehren96}
\bysame, \emph{Weak {C}$^*$ {H}opf symmetry}, Group Theoretical Methods in
  Physics (A.~Bohm, H.-D. Doebner, and P.~Kielanowski, eds.), Lecture Notes in
  Physics, vol. 504, Heron Press, Berlin, 1997, q-alg/9611007, pp.~62--69.

\bibitem{Roberts76}
J.E. Roberts, \emph{Local cohomology and superselection structure}, Commun.\
  Math.\ Phys. \textbf{51} (1976), 107--119.

\bibitem{Roberts80}
\bysame, \emph{Net cohomology and its applications to field theory}, Quantum
  Fields -- Algebras, Processes (L.~Streit, ed.), Springer, Wien, New York,
  1980, pp.~239--268.

\bibitem{Roberts}
\bysame, \emph{Lectures on algebraic quantum field theory}, The Algebraic
  Theory of Superselection Sectors. Introduction and Recent Results
  (D.~Kastler, ed.), World Scientific, Singapore, New Jersey, London, Hong
  Kong, 1990, pp.~1--112.

\bibitem{S94}
B.~Schroer, \emph{Modular theory and symmetry in {QFT}}, Mathematical Physics
  towards the $21^{st}$ Century (R.N. Sen and A.~Gersten, eds.), Ben-Gurion of
  the Negev Press, Beer-Sheva, Israel, 1994.

\bibitem{SHW98}
B.~Schroer and H.-W. Wiesbrock, \emph{Modular theory and geometry}, Rev. Math.
  Phys. \textbf{12} (2000), 139--158.

\bibitem{St}
O.~Steinmann, \emph{A {Jost-Schroer} theorem for string fields}, Commun. Math.
  Phys. \textbf{87} (1982), 259--264.

\bibitem{Stratila}
S.~Str$\check{\rm a}$til$\check{\rm a}$, \emph{Modular theory in operator
  algebras}, Abacus Press, Tunbridge Wells, England, 1981.

\bibitem{Unruh}
W.G. Unruh, \emph{Notes on black hole evaporation}, Rev. Math. Phys.
  \textbf{14} (1976), 870--892.

\bibitem{Var2}
V.S. Varadarajan, \emph{Geometry of quantum theory}, vol.~{II}, Van Nostrand
  Reinhold Co., New York, 1970.

\bibitem{Wil}
F.~Wilczek, \emph{Quantum mechanics of fractional-spin particles}, Phys.\ Rev.\
  Lett. \textbf{49} (1982), 957--1149.

\end{thebibliography}
\end{document}